\documentclass[english,reprint, longbibliography, superscriptaddress, breaklinks=true, showkeys, showpacs=false, nofootinbib]{revtex4}
\usepackage[T1]{fontenc}
\usepackage[latin9]{inputenc}
\usepackage{color}
\usepackage{babel}
\usepackage{amsmath}
\usepackage{amsthm}
\usepackage{amsfonts}
\usepackage{amssymb}
\usepackage{graphicx}
\usepackage{subfigure}
\usepackage{physics}

\usepackage[unicode=true,pdfusetitle,
 bookmarks=true,bookmarksnumbered=false,bookmarksopen=false,
 breaklinks=true,pdfborder={0 0 0},backref=false,colorlinks=true]
 {hyperref} 
\usepackage{breakurl}

\makeatletter
\@ifundefined{textcolor}{}
{%
 \definecolor{BLACK}{gray}{0}
 \definecolor{WHITE}{gray}{1}
 \definecolor{RED}{rgb}{1,0,0}
 \definecolor{GREEN}{rgb}{0,1,0}
 \definecolor{BLUE}{rgb}{0,0,1}
 \definecolor{CYAN}{cmyk}{1,0,0,0}
 \definecolor{MAGENTA}{cmyk}{0,1,0,0}
 \definecolor{YELLOW}{cmyk}{0,0,1,0}
}

\pdfoutput=1
\hypersetup{colorlinks=true,citecolor=blue,linkcolor=cyan,urlcolor=blue,filecolor= green, breaklinks=true}
\usepackage{url}
\usepackage{breakurl}

\makeatother

\begin{document}

\title{Complete complementarity relations in system-environment decoherent dynamics}

\author{Marcos L. W. Basso}
\email{marcoslwbasso@mail.ufsm.br}
\address{Departamento de F\'isica, Centro de Ci\^encias Naturais e Exatas, Universidade Federal de Santa Maria, Avenida Roraima 1000, Santa Maria, Rio Grande do Sul, 97105-900, Brazil}

\author{Jonas Maziero}
\email{jonas.maziero@ufsm.br}
\address{Departamento de F\'isica, Centro de Ci\^encias Naturais e Exatas, Universidade Federal de Santa Maria, Avenida Roraima 1000, Santa Maria, Rio Grande do Sul, 97105-900, Brazil}

\selectlanguage{english}%

\begin{abstract}
\textbf{Abstract}
We investigate the system-environment information flow from the point of view of complete complementarity relations. We consider some commonly used noisy quantum channels: Amplitude damping, phase damping, bit flip, bit-phase flip, phase flip, depolarizing, and correlated amplitude damping.
By starting with an entangled bipartite pure quantum state, with the linear entropy being the quantifier of entanglement, we study how entanglement is redistributed and turned into general correlations between the degrees of freedom of the whole system. For instance, it is possible to express the entanglement entropy in terms of the multipartite quantum coherence or in terms of the correlated quantum coherence of the different partitions of the system. In addition, we notice that for the depolarizing and bit-phase flip channels the wave and particle aspects can decrease or increase together. Besides, by considering the environment as part of a pure quantum system, the linear entropy is shown to be not just a measure of mixedness of a particular subsystem, but a correlation measure of the subsystem with rest of the world.
\end{abstract}


\maketitle

\section{Introduction}
With the development of quantum information science and the increased interest in quantum foundations, together with the fact that the wave-particle duality was one of the cornerstones in the development of quantum mechanics, recently this intriguing quantum feature has been receiving a lot of attention among researchers. Such aspect is generally captured, in a qualitative way, by Bohr's complementarity principle \cite{Bohr}. It states that quantons \cite{Leblond} have characteristics that are equally real, but mutually exclusive. The wave aspect is characterized by interference fringes' visibility, or, more generally, by coherent superposition of the eigenbasis of some observable, meanwhile the particle nature is given by the which-way information of the path along the interferometer, or, more generally, through our knowledge of the particular eigenstate of the chosen observable. In principle, the complete knowledge of the path destroys the interference pattern visibility and vice-versa. However, in the quantitative scenario explored by Wootters and Zurek \cite{Wootters}, they showed that simultaneous-partial path information and interference pattern visibility can be retained. Later, this work was extended by Englert, who derived a wave-particle duality relation \cite{Engle}. On the other hand, using a different line of reasoning, Greenberger and Yasin \cite{Yasin} considered a two-beam interferometer, in which the intensity of each beam was not necessarily the same, and defined a measure of path information called predictability. In this scenario, if the quantum system passing through the beam-splitter has different probability of getting reflected in the two paths, one could have some path information of the quantum system, which resulted in a different kind of wave-particle duality relation
\begin{equation}
    P^2 + V^2 \le 1 \label{eq:cr1},
\end{equation}
where $P$ is the predictability and $V$ is the visibility of the interference pattern. Hence, by examining Eq. (\ref{eq:cr1}) one can see that an experiment can also provide partial information about the wave and particle nature of the quanton. For instance, in Ref. \cite{Auccaise} the authors confirmed that it is possible to measure both aspects of the system with the same experimental apparatus. As aforementioned, this topic received a lot of attention among researchers, which lead to the quantification of Bohr's complementarity principle, and D\"urr \cite{Durr}  and Englert et al. \cite{Englert} established minimal and reasonable conditions that any visibility and predictability measure should satisfy. As well, it was suggested that the quantum coherence \cite{Baumgratz} would be a good generalization of the visibility measure \cite{Bera, Bagan, Tabish, Mishra}. Until now, many approaches were taken to obtain complementarity relations like the one in Eq. (\ref{eq:cr1}) \cite{Angelo, Coles, Hillery, Qureshi, Maziero}. Among them, it is worth mentioning that Baumgratz et al., in Ref. \cite{Baumgratz}, showed that the $l_1$-norm and the relative entropy of coherence are bone-fide measures of coherence, meanwhile the Hilbert-Schmidt (or $l_2$-norm) coherence is not. However, as shown in Ref. \cite{Maziero}, all of these measures of quantum coherence, including the Wigner-Yanase quantum coherence \cite{yu_Cwy}, are bone-fide measures of visibility.

On the other hand, as emphasized by Qian et al. \cite{Qian}, complementarity relations like Eq. (\ref{eq:cr1}) do not really express a balanced exchange between $P$ and $V$, once the inequality permits a decrease of $P$ and $V$ together, or an increase by both, although both measures remain complementary to each other. It even allows the extreme case $P = V = 0$ to occur, for which neither the ``waveness'' or ``particleness'' of the quanton is observed. Thus, no information about the quanton is obtained while, in an experimental setup, we still have a quanton on hands. Therefore, one can see that something must be missing from Eq. (\ref{eq:cr1}). As noticed by Jakob and Bergou \cite{Janos}, this lack of information about the system is due to another unique quantum feature: entanglement \cite{Bruss, Horodecki}. This means that the information is being shared with another system and this kind of quantum correlation can be seen as responsible for the loss of purity of each subsystem such that, for pure maximally entangled states, it is not possible to obtain information about the local properties of the subsystems. Therefore, to fully characterize a quanton it is not enough to consider its wave-particle properties; one has also to regard its correlations with other subsystems. In addition, Qian et al. provided the first experimental confirmation of the complete complementarity relation (CCR) using single photon states.  Lastly, as we reported in Refs. \cite{Marcos, Basso}, for each coherence and predictability measure that quantifies the local properties of a quanton, there is a corresponding correlation measure that completes each complementarity relation, if we consider the quanton as part of a multipartite pure quantum system.

In particular, in Ref. \cite{Marcos} we considered an $n$-quanton pure state described by $\ket{\Psi}_{A_1,...,A_n} \in \mathcal{H}_{1} \otimes ... \otimes \mathcal{H}_{n}$ with dimension $d = d_{A_1}d_{A_2}...d_{A_n}$. By defining a local orthonormal basis for each subsystem $A_m$, $\{\ket{j_m}_{A_m}\}_{j = 0}^{d_{A_m} - 1}$ with $m = 1,...,n$, in the local Hilbert spaces $\{\mathcal{H}_m\}_{m = 1}^{n}$, the subsystem $A := A_1$ can be represented by the reduced density operator $\rho_A = \Tr_{A_2,...,A_n} (\ket{\Psi}_{A_1,...,A_n} \bra{\Psi})$ \cite{Mark}. By exploring the expression of the purity of the multipartite density matrix, $1 - \Tr \rho^2_{A_1 A_2 ... A_n} = 0$, we derived the complete complementarity relation \cite{Marcos}:
\begin{equation}
    P_{hs}(\rho_A) + C_{hs}(\rho_A) + S_l(\rho_A) = (d_A - 1)/d_A, \label{eq:ccrhs},
\end{equation}
where 
\begin{equation}
C_{hs}(\rho_A) = \sum_{j \neq k} \abs{\rho^A_{jk}}^2
\end{equation}
is the Hilbert-Schmidt quantum coherence \cite{Jonas1} and \begin{equation}
P_{hs}(\rho_A) := S^{\max} - S_l(\rho_{diag}) =  \sum_j (\rho^A_{jj})^2 - 1/d_A
\end{equation}
is the corresponding predictability measure, where $\rho_{diag}$ is the diagonal elements of $\rho$, and 
\begin{equation}
S_l(\rho_A) := 1 - \Tr \rho_A^2
\end{equation}
is the linear entropy, which altogether quantifies the behaviour of the quanton $A$. It is noteworthy that in Ref. \cite{Baumgratz} the authors established minimum conditions that any measure of coherence must satisfy. However, it is worth to emphasize that the criteria for measures of coherence are not the same as the criteria for measures of visibility \cite{Durr, Englert}. So much so that Hilbert-Schmidt's (or $l_2$-norm) quantum coherence is considered a good measure of visibility, as shown in Ref. \cite{Maziero}, while it does not meet all the criteria for a good measure of quantum coherence, since it does not satisfy the condition of not increasing under incoherent operations. Regarding the intuition about the measure of predictability, let us consider the projector onto the state index $j$: $\Pi_j := \ketbra{j}$. Such state can be the one of the paths of a Mach-Zehnder interferometer. Now, the uncertainty (the variance) of the path $j$ is given by
\begin{equation}
    \mathcal{V}(\rho,\Pi_j) 
    = \expval{\Pi^2_j} - \expval{\Pi_j}^2  = \Tr \rho \Pi_j^2 - (\Tr \rho \Pi_j)^2 = \rho_{jj} - \rho^2_{jj},
\end{equation} 
such that the uncertainty of all paths is given by sum over $j$
\begin{equation}
    \sum_j \mathcal{V}(\rho,\Pi_j) = 1 - \sum_j \rho^2_{jj},
\end{equation}
which is exactly the linear entropy of $\rho_{diag}$, $S_l(\rho_{diag})$. Thus, after repeating the same experiment several times, we obtain a probability distribution given by $\rho_{00},..., \rho_{d-1 d-1},$ which represents a probability of the quanton being measured in the state $\ket{0},..., \ket{d - 1}$. From this probability distribution, it is possible to calculate the uncertainty about the paths through $S_l(\rho_{diag})$ such that $P_{hs}(\rho) := S_{l}^{max} - S_{l}(\iota_{\rho}^{hs})$ offers a measure of the capability to predict what outcome will be obtained the next time we run the experiment. For instance, if after repeating the experiment several times, we obtain a uniform  probability distribution, i.e., $\{\rho_{jj}= 1 /d\}_{j = 0}^{d-1}$, our ability to make a prediction is null. Now, if $\{\rho_{jj}\}_{j = 0}^{d-1}$ does not represent a uniform probability distribution, then $P_{hs}(\rho) \neq 0$. Therefore, it is possible to see that $P_{hs}(\rho) = \sum_{j} \rho^2_{jj} - 1/d$ is a way of quantifying how much the probability distribution $\{\rho_{jj}\}_{j = 0}^{d-1}$ differs from the uniform probability distribution.

It is worthwhile mentioning that the CCR in Eq. (\ref{eq:ccrhs}) is a natural generalization of the complementarity relation obtained by Jakob and Bergou \cite{Jakob, Bergou} for bipartite pure quantum systems. Moreover, in Ref. \cite{Bhaskara}, it was found out that, for a general multipartite quantum system, $S_l(\rho_A)$ is related to the generalized concurrence. Also, if the subsystem $A$ is not correlated with rest of subsystems, then $A$ is pure and $S_l(\rho_A) = 0$. However, no system is completely isolated from its environment, unless, perhaps, the universe as a whole. This interaction between the system and the environment causes them to correlate, leading to an irreversible transfer of information from the system to the environment. This process, called decoherence, results in a non-unitary dynamics for the system, whose most important effect is the disappearance of phase relationships between the subspaces of the system Hilbert space \cite{Zurek}, being responsible for the emergence of classical behaviour from the quantum realm \cite{Zurek1}. Therefore, the interaction between the quantum system and the environment can be seen responsible for the mixedness of the quantum system. In this case, if we look only to the $n$-partite mixed quantum systems described by $\rho_{A_1,...,A_n}$, then $S_l(\rho_{A}) = 1 - \Tr \rho^2_{A}$ measures not only quantum correlations, but, more generally, the noise introduced by correlations with the environment. Then, we can interpret the entropy functional $S_l(\rho_{A})$ as the mixedness of the subsystem $A$, as defined in Ref. \cite{Dhar}, and we still have a complete complementarity relation, but it can not be derived from the purity of the multipartite density matrix of the system alone. 

In this article, we consider the environment as part of the quantum system such that the whole system can be taken as a multipartite pure quantum system, which allows us to consider $S_l(\rho_A)$ not just as a measure of mixedness of the subsystem $A$, but as a correlation measure of the subsystem $A$ with rest of the world. The main purpose of this work is to show that it is possible to keep track of how the correlations are redistributed among the different partitions of the system and how this affects the complementarity behavior of a quanton. Therefore, we study the behavior of CCR under the action of the most common noisy quantum channels (amplitude damping, phase damping, bit flip, phase flip, bit-phase flip, depolarizing, and correlated amplitude damping), since, until now, no systematic study about this subject was done. Hence, this work takes the first step in this direction. In addition, by starting with an entangled bipartite pure quantum state, with the linear entropy being the quantifier of entanglement, we study how the entanglement entropy is redistributed and turned into general correlations between the degrees of freedom of the whole system. For instance, as we will show, it is always possible to express the entanglement entropy in terms of the multipartite quantum coherence \cite{Yao}, or in terms of the correlated quantum coherence \cite{Tan}, of the different partitions of the system, that in some cases is related with entanglement but in other cases is not. It is important to emphasize that although the formalism of noisy quantum channels does not mention the environment's nature, it is mathematically and physically useful and has been applied in several previous works \cite{Salles, Wang, Jonas}.

The reminder of this article is organized as follows. In Sec. \ref{sec:comp} we give a brief review of the dynamics of open quantum systems, by mentioning the Kraus' operator-sum representation and its corresponding unitary mapping. Also, we show that it is possible to consider the environment as part of the quantum system such that the whole system can be taken as a multipartite pure quantum system. Afterwards, we apply this formalism to study the behavior of CCRs and the dynamics of the quantum correlations of two entangled qubits evolving under the action of environments modeled by the amplitude damping (Sec. \ref{subsec:ampli}), correlated amplitude damping (Sec. \ref{subsec:correamp}), phase damping (Sec. \ref{subsec:phasedam}), bit flip (Sec. \ref{subsec:bitplip}), phase flip (Sec. \ref{subsec:phaseflip}), bit-phase flip (Sec. \ref{subsec:bitphase}), and the depolarizing (Sec. \ref{subsec:depo}) channels. We present our conclusions in Sec. \ref{sec:conc}.

\section{Dynamics of complementarity under quantum channels}
\label{sec:comp}
The time evolution of quantum systems is represented by unitary maps $\ket{\Psi} \to U \ket{\Psi}$, or equivalently, $\rho \to U \rho U^{\dagger}$, where $U$ is a unitary operator. However, this is not the most general evolution, since it is always possible to couple the quantum system with another one (like the environment), evolve both with a unitary operator that can create entanglement between the two systems, and then ignore the second system \cite{Mark}. Hence, a natural way to define the dynamics of an open quantum system is to consider the unitary evolution of the global system constituted by the system of interest $S$ and its environment $E$ and ignore (take the partial trace over) the environment variables \cite{Breuer}:
\begin{equation}
    \varepsilon(\rho_S) = \Tr_E (U (\rho_{S} \otimes \rho_E) U^{\dagger}), 
\end{equation}
where $U$ is the unitary operator that describes the evolution of the global system, meanwhile $\rho_{S} \otimes \rho_E$ is the initial state of the whole system, which is uncorrelated. Without loss of generality, $\rho_E$ can be chosen as a pure state $\rho_{E} = \ket{0}_E\bra{0}$, due to the purification theorem \cite{nielsen}. For instance, $\ket{0}_E$ can represent the vacuum modes of the electromagnetic field. The resulting evolution for the system $S$ when ignoring the degrees of freedom of the environment $E$, $\varepsilon(\rho_S)$, will be, in general, non-unitary and can be described by a non-unitary quantum map: $\varepsilon: \rho_S \to \varepsilon(\rho_S)$, which is a completely positive and trace preserving map \cite{nielsen}. Hence, the map $\varepsilon$ can be written in the so called operator-sum representation
\begin{equation}
    \varepsilon(\rho_S) = \sum_i K_i \rho_S K^{\dagger}_i,
\end{equation}
where $K_i := \bra{i}U\ket{0}_E$, the Kraus' operators, act only
on the Hilbert space of system $S$, meanwhile the set $\{\ket{i}_E\}$ is an orthonormal
basis for the environment. The Kraus' operators satisfy $\sum_i K^{\dagger}_i K_i = I_{\mathcal{H}_S}$, yielding a map $\varepsilon$ that is
trace-preserving, with $I_{\mathcal{H}_S}$ being the identity in the Hilbert space of the system $S$. It is worthwhile mentioning that the set of Kraus' operators inducing a certain evolution on the system state is not unique, with $\{K_i\}$ and $\{K'_i = \sum_l V_{il} K_l \}$ leading to the same $\varepsilon(\rho_S)$, where $V$ is a unitary transformation.

Since, in this article, we are interested in bipartite pure quantum systems $S := AB$, it is necessary to specify which type of environment we are dealing with. For instance, it is possible to consider two types of environments: (i) local and (ii) global. In case (i), each part of the system $S$ interacts with its local-independent environment. Hence, correlations between the parts of the system cannot be increased by interaction with the environment. For instance, memoryless quantum channels describe those scenarios in which the noise acts identically and independently on each subsystem \cite{Caruso}. Thus, the dynamics of the joint state of the subsystems $A$ and $B$ is given by the following expression
\begin{equation}
    \varepsilon(\rho_{AB}) = \sum_{i,j}(K^A_i \otimes K^B_j) \rho_{AB} (K^A_i \otimes K^B_j)^{\dagger}, \label{eq:kraus}
\end{equation}
where $\{K^A_i\}$, $\{K^B_j\}$ are the Kraus' operators acting in the Hilbert space of the subsystems $A$ and $B$, respectively. Whereas, in case (ii) the interaction of all parts of $S$ with the same environment may lead, in principle, to an increase of the correlations between the parts of the system due to non-local interactions mediated by the environment \cite{An}. Also, in this case, whenever  the  tensorial  decomposition of the Kraus operators in Eq. (\ref{eq:kraus}) does not apply, one can speak of memory channels or correlated noise channels, as e.g. the correlated amplitude damping channel \cite{Falci}.

Now, if $\{\ket{j}_A \otimes \ket{k}_B := \ket{j,k}_{A,B}\}_{j,k = 0}^{d_A - 1, d_B - 1}$  is a complete basis for the bipartite quantum system $AB$, there exists a Kraus' representation involving up to $(d_A d_B)^2$ Kraus' operators. Since we shall deal with the most common memoryless quantum channels and the correlated amplitude damping channel, together with the fact that the initial state of the bipartite quantum system is pure and described by $\ket{\Psi}_{A,B} \in \mathcal{H_{AB}}$, the dynamics of the joint state of the system and the environment can be represented by the following map
\begin{equation}
    U \ket{\Psi}_{A,B} \otimes \ket{0,0}_{E_A,E_B} = \sum_{i,j} (K_{ij} \ket{\Psi}_{A,B}) \otimes \ket{i,j}_{E_A, E_B},
\end{equation}
where $K_{ij} = K^A_i \otimes K^B_j$ for the memoryless quantum channels.

It is important to recognize that, besides the quantum channel representation of system-environment interactions being appealing from the experimental point of view, one can obtain the Kraus' operators describing the noise channel acting on the system by applying a phenomenological approach \cite{lucas} or by using quantum process tomography \cite{chuang}, and then one can use this information to investigate e.g. the system-environment quantum properties, without much concern about the usually complicated structure of the environment.

Now, if we consider the states of the environment as part of our system, we have a multipartite pure quantum system, which allows us to derive complete complementarity relations for the subsystem $A$ (or for any other subsystem) with $S_l(\rho_A)$ measuring the entanglement of $A$ with the rest of the quantum system. Once $\rho_{A B E_A E_B} = \ket{\Psi}_{A,B}\bra{\Psi} \otimes \ket{0,0}_{E_A,E_B}\bra{0,0}$ and $\rho'_{A B E_A E_B} = U \rho_{A B E_A E_B} U^{\dagger}$, it follows that $\Tr (\rho'_{A B E_A E_B})^2 = \Tr (\rho_{A B E_A E_B})^2$, and the purity of the whole system is preserved. Therefore, it is possible to derive Eq. (\ref{eq:ccrhs}) from the purity of the entire system and to take $S_l(\rho_A)$ as a measure of correlations between $A$ and the rest of the system.
    
\subsection{Amplitude damping channel}
\label{subsec:ampli}

The (memoryless) amplitude damping channel (ADC) is a schematic model of the decay of an excited state of a (two-level) atom due to spontaneous emission of a photon. There is an exchange of energy between system and the environment, such that the system is driven into thermal equilibrium with the environment. This channel may be modeled by treating $E$ as a large collection of independent harmonic oscillators interacting weakly with $S$. We denote the ground state of the system as $\ket{0}_k$ and the excited state as $\ket{1}_k$ for $k = A,B$.  The environment is the electromagnetic field, assumed initially to be in its vacuum state $\ket{0}_{E_k}$, $k = A,B$.  After the coupling between the quanton and the environment, there is a probability $p \in [0,1]$ that the excited state of the quanton has decayed to the ground state and a photon has been emitted, so that the environment has made a transition from the "no photon" state $\ket{0}_{E_k}$ to the "one photon" state $\ket{1}_{E_k}$ \cite{Knight}.  This evolution is described by a unitary transformation that acts on the quanton and its environment according to
\begin{align}
    U \ket{j,0}_{k, E_k} := \delta_{j,0}\ket{0,0}_{k,E_k} + \delta_{j,1}(\sqrt{1 - p} \ket{1,0}_{k,E_k} + \sqrt{p} \ket{0,1}_{k,E_k}),
\end{align}
where $j = 0,1$ and $k = A,B$. In this case, each subsystem has its own environment and the probability of each one of them decaying is the same. 

We consider the initial state of the bipartite quantum system $AB$ to be \begin{equation}
\ket{\Psi}_{A,B} = x \ket{0,0}_{A,B} + \sqrt{1 - x^2}\ket{1,1}_{A,B},
\end{equation}
with $x \in [0,1]$. Then, initially, the predictability of the quanton A is given by $P_{hs}(\rho_A)_{p=0} = 1/2 - 2x^2(1 - x^2)$, and the entanglement entropy $S_l(\rho_A)_{p = 0} = C^c_{hs}(\rho_{AB})_{p = 0} = \frac{1}{2}\mathcal{C}^2_{AB} = 2x^2(1 - x^2)$, where $C^c_{hs}$ is the \textit{correlated quantum coherence} \cite{Tan}, which, in this case, is equal to the bipartite quantum coherence, and is defined as 
\begin{equation}
C^c(\rho_{AB}) := C(\rho_{AB}) - C(\rho_{A}) - C(\rho_{B})
\end{equation}
for some measure of coherence $C$, since in Ref. \cite{Tan} the authors shows that such a measure can be considered asa measure of quantum correlations as defined by the asymmetric and symmetrized versions of quantum discordas well as quantum entanglement, respectively, thus providing a unified view of these correlations. While $\mathcal{C}_{AB}$ is the concurrence measure of entanglement \cite{Woot}, which, for pure states, is defined as $\mathcal{C}_{AB} := \sqrt{2(1 - \Tr \rho^2_{A(B)})}$. For X states, there is a simple closed expression  given by \cite{Quesada}: $\mathcal{C}_{AB} = 2 \max(0, \Lambda_1, \Lambda_2)$, with $\Lambda_1 = \abs{\rho_{14}} - \sqrt{\rho_{22}\rho_{33}}$ and $\Lambda_2 = \abs{\rho_{23}} - \sqrt{\rho_{11}\rho_{44}}$. After the coupling with the environment, the evolved global state is a pure state given by

\begin{figure}[t]
\subfigure[\footnotesize Predictability and Correlation measures as a function of $p$.]{\includegraphics[scale = 0.5]{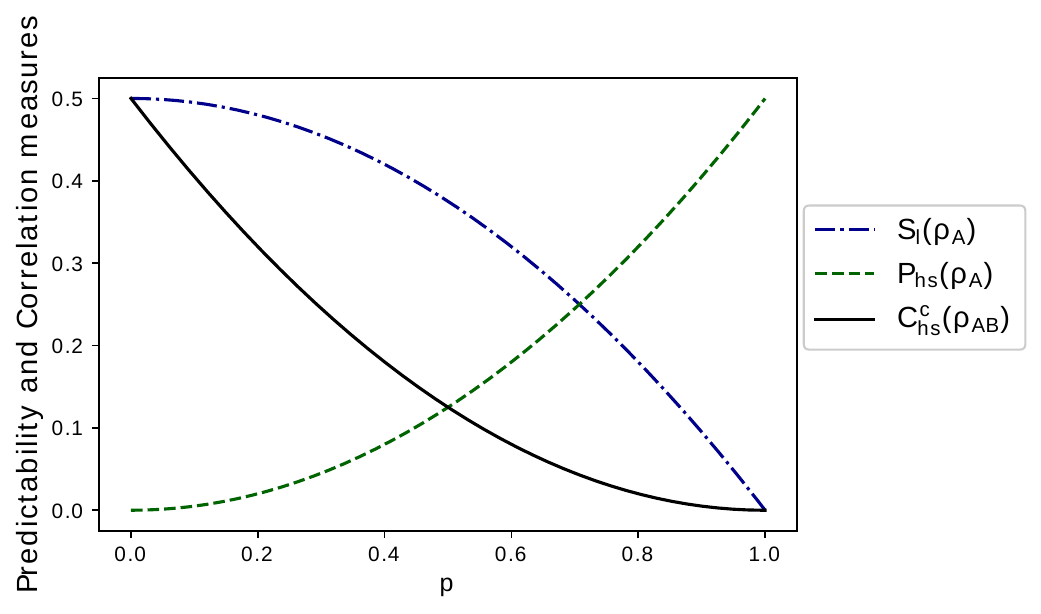}\label{fig:ccc}}
\subfigure[\footnotesize  Predictability $+$ Correlation as a function of $p$.]{\includegraphics[scale = 0.5]{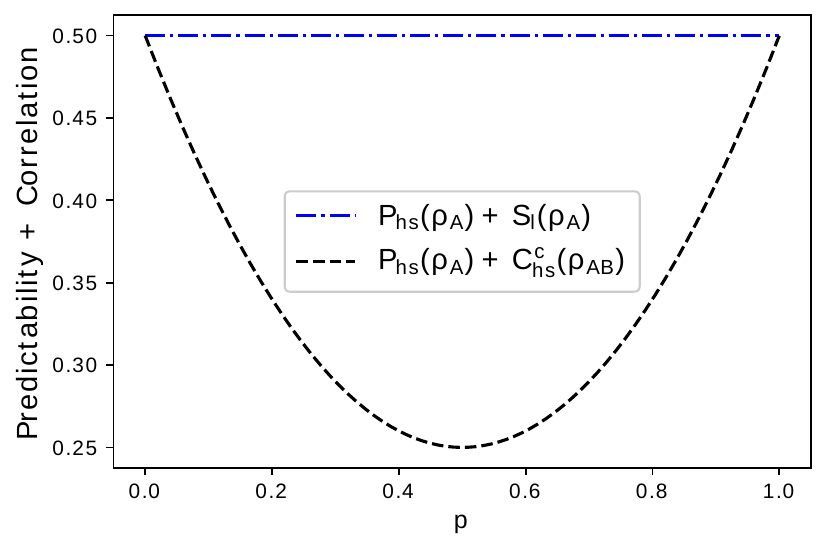}\label{fig:cc} }
\caption{The different aspects of the quanton $A$ as a function of $p$ for the amplitude damping channel; for the initial state with $x=1/\sqrt{2}$.}
\label{fig:examp}
\end{figure}

\begin{align}
    \ket{\Psi}_{A,B,E_A,E_B} & := U \ket{\Psi}_{A,B} \otimes \ket{0,0}_{E_A,E_B}\\
    & = x\ket{0,0}\ket{0,0} + \sqrt{1 - x^2}\Big((1-p)\ket{1,1}\ket{0,0} + \sqrt{p(1 - p)}(\ket{1,0}\ket{0,1}\nonumber \\ & + \ket{0,1}\ket{1,0}) + p \ket{0,0}\ket{1,1}\Big),
\end{align}
which, in general, is an entangled state between all degrees of freedom. The reduced density operator for the partition $AB$, obtained by taking the partial trace
of $\rho_{A B E_A E_B}$  over the degrees of freedom of the reservoir, i.e, $\rho_{AB} = Tr_{E_A E_B}(\rho_{A B E_A E_B})$, is given by
\begin{equation}
    \rho_{AB} =  \begin{pmatrix}
x^2 + p^2(1 - x^2) & 0 & 0 & (1- p) x\sqrt{1 - x^2}\\
0 & p(1-p)(1 - x^2) & 0 & 0\\
0 & 0 & p(1-p)(1 - x^2) & 0\\
(1- p) x\sqrt{1 - x^2} & 0 & 0 & (1 - p)^2(1 - x^2)\\
\end{pmatrix},
\end{equation}
which, in general, is no longer a pure state, and the off diagonal elements are affected by a factor of $1 - p$. It is interesting to notice that the previsibility of the subsystem $A$ is also affected, since $\rho_A = (x^2 + p(1 - x^2))\ketbra{0} + (1 - p)(1 - x^2)\ketbra{1}$. For $x \ge \sqrt{1 - x^2}$, $\rho_{AB}$ is entangled  $\forall p \in [0,1)$, by the Peres' separability (PPT) criterion \cite{Asher}, and $\mathcal{C}_{AB} > 0$ for all $p \in [0,1)$, vanishing only at $p = 1$. In particular, taking $x = 1/\sqrt{2}$, one can see that the subsystems are also entangled with the environment, since $P_{hs}(\rho_A) = \frac{1}{2}p$, $C^c_{hs}(\rho_{A,B}) = \frac{1}{2} \mathcal{C}_{A,B} = \frac{1}{2}(1 - p)^2$ and $S_l(\rho_A) = \frac{1}{2}(1 - p)$. For $p = 0$ (no coupling between the bipartite quantum system and the environment), the predicatibility is zero, meanwhile the entanglement between $A$ and $B$ is maximum, and $S_l(\rho_A) = C^c_{hs}(\rho_{AB})$. However, as the global state evolves, we shall have $S_l(\rho_A) > C^c_{hs}(\rho_{AB})$, Fig. \ref{fig:ccc}, indicating that $S_l(\rho_A)$ is measuring the correlations of $A$ with the whole system, and therefore $C^c_{hs}(\rho_{AB})$ is not enough to completely quantify the correlations of $A$ with rest of the system, as represented in Fig. \ref{fig:cc}.

\begin{figure}[t]
\subfigure[\footnotesize Correlation measures as a function of $p$ for the state $x = 1/4$.]{\includegraphics[scale = 0.5]{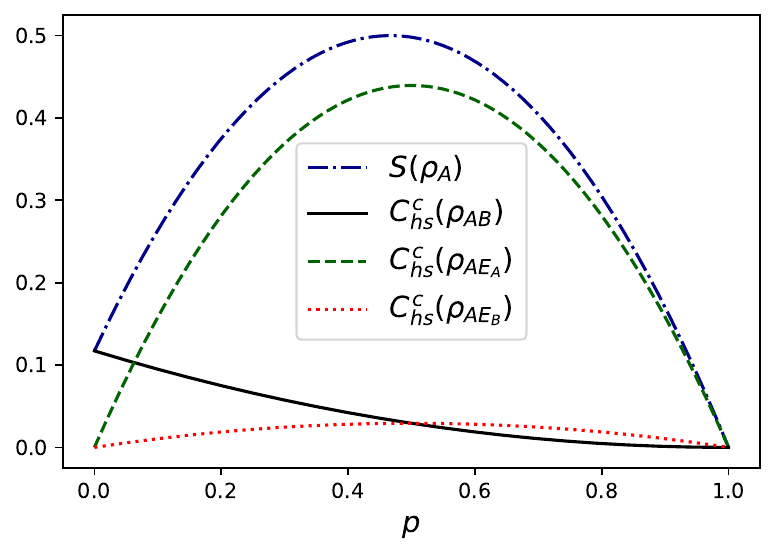}\label{fig:amp}}
\subfigure[\footnotesize Correlation measures as a function of $p$ for the state $x = 1/\sqrt{2}$.]{\includegraphics[scale = 0.5]{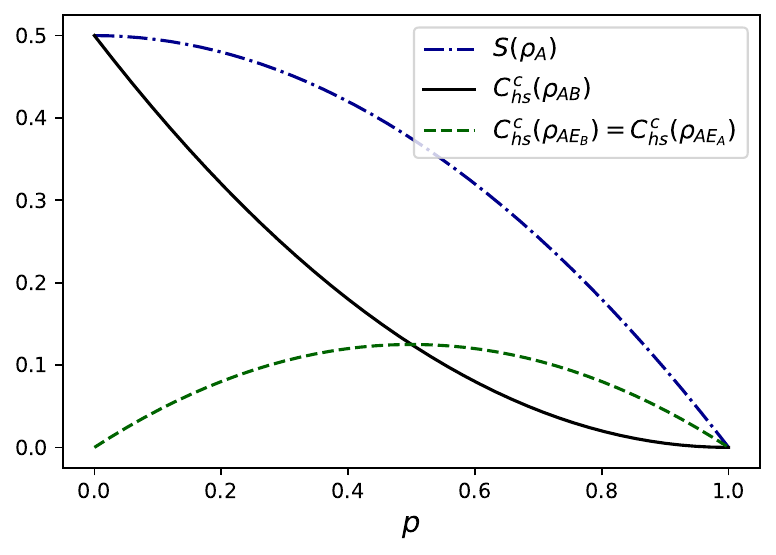}\label{fig:amp2}}
\subfigure[\footnotesize Comparison between $C^c_{re}$ and $C^c_{hs}$ as a function of $p$ for $x = 1/\sqrt{2}$.]{\includegraphics[scale = 0.5]{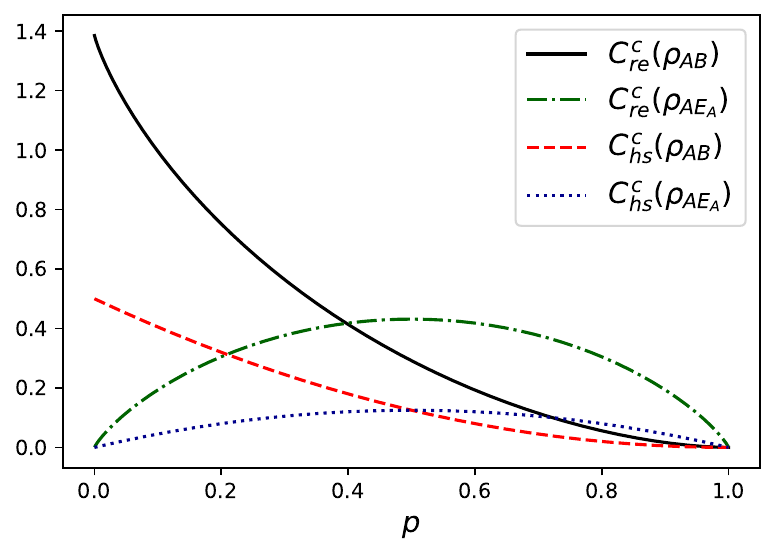}\label{fig:coecorr}}
\subfigure[\footnotesize Correlation measures as a function of $p$ for $x = 0.5$. ]{\includegraphics[scale = 0.5]{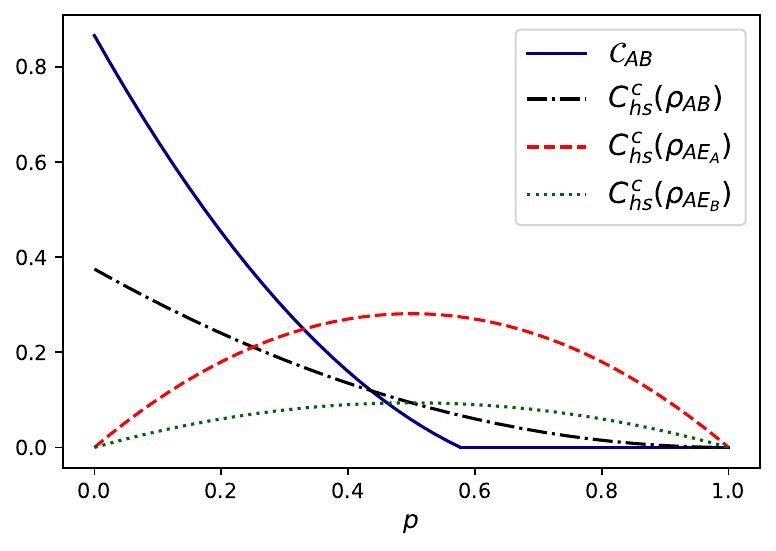}\label{fig:sudden}}
\caption{Entanglement entropy of $A$ and correlated quantum coherence of the different bipartitions as a function of $p$ for the amplitude damping channel.}
\label{fig:examp1}
\end{figure}

On the other hand, for $x < \sqrt{1 - x^2}$, the entanglement between $A$ and $B$ goes to zero at $p = x/\sqrt{1 - x^2}$, which is known as the entanglement sudden death \cite{Eberly, Melo}. If the parametrization $1 - p = e ^{-\Gamma t}$ \cite{Salles} is used, this implies a finite-time disentanglement, even though the bipartite coherence, which is equal to the correlated coherence, goes to zero only asymptotically, as represented in Fig. \ref{fig:sudden}. We can see that the correlated coherence of $AB$ goes to zero asymptotically, whereas the correlated coherence of $A$ and the environments $E_A$ and $E_B$ increases until $p/2$, as we shall see below. By looking at the reduced states of $A$ with the environments $E_A$ and $E_B$,
\begin{align}
        & \rho_{AE_A} =  \begin{pmatrix}
x^2  & 0 & 0 & 0\\
0 & p(1 - x^2) & \sqrt{p(1-p)}(1 - x^2)  & 0\\
0 & \sqrt{p(1-p)}(1 - x^2)  & (1-p)(1 - x^2)  & 0\\
0 & 0 & 0 & 0\\
\end{pmatrix}, 
\end{align}

\begin{align}
& \rho_{AE_B} =  \begin{pmatrix}
x^2 + p(1 - p)(1 - x^2) & 0 & 0 & \sqrt{p(1- p)} x\sqrt{1 - x^2}\\
0 & p^2(1 - x^2) & 0 & 0\\
0 & 0 & (1-p)^2(1 - x^2) & 0\\
\sqrt{p(1- p)} x\sqrt{1 - x^2} & 0 & 0 & p(1 - p)(1 - x^2)\\
\end{pmatrix},
\end{align}
we see that the correlated quantum coherence is given by $C^{c}_{hs}(\rho_{AE_A}) = 2(1-x^2)^2p(1-p)$ and $C^{c}_{hs}(\rho_{AE_B}) = 2x^2(1 - x^2)p(1-p)$.  Therefore 
\begin{equation}
S_l(\rho_A) = C^{c}_{hs}(\rho_{AB}) +C^{c}_{hs}(\rho_{AE_A})  + C^{c}_{hs}(\rho_{AE_B}).
\end{equation}
This leads to the following complete complementarity relation
\begin{align}
  P_{hs}(\rho_A) +  C^{c}_{hs}(\rho_{AB}) +C^{c}_{hs}(\rho_{AE_A})  + C^{c}_{hs}(\rho_{AE_B}) = 1/2.
\end{align}
It is interesting noticing that, for the states characterized by $x = 0.25$ and $0.5$ (see Figs. \ref{fig:amp} and \ref{fig:sudden}, respectively), the correlations, as a whole, increase until they reach their maximum possible values even though the entanglement between $A$ and $B$ decreases monotonically going to zero in a finite time. When $S_l(\rho_A)$ is maximum, the predictability is zero, and the information of the quanton $A$ is completely diluted in the correlations with $B, E_A$ and $E_B$. On the other hand, for the state given by $x = 1/\sqrt{2}$ (initially maximally entangled with $B$), in Fig. \ref{fig:amp2}, the correlations as whole decrease such that, when the predictability is zero, the information of the quanton $A$ is completely shared with the system $B$. Also, it is worth pointing out that, due to the inherent symmetry of the system, the same results are valid for the subsystem $B$. In addition, for the particular case $x = 1/ \sqrt{2}$, we can see that $A$ is entangled with $E_A$ and $E_B$, $\forall p \in (0,1)$, by the Peres' separability criterion, but just the correlated quantum coherence of $AE_A$ is directly related to the concurrence measure by the expression $C^c_{hs}(\rho_{AE_A}) = \frac{1}{2}\mathcal{C}^2_{AE_A}$. Besides, it's worth pointing out that the measures of correlated coherence of the bipartition $A$ and the environment $E_A$ ($E_B$) is invariant under local unitary transformations on $E_A$ ($E_B$), if the local quantum coherence of the environment remains zero under such transformation \cite{Tann}. In Fig. \ref{fig:coecorr}, we compare $C_{hs}^c$ and $C_{re}^c$, where $C_{re}^c$ is the basis-independent relative entropy of correlated coherence, which is equal to the quantum mutual information \cite{Leopoldo}. As one can see, the qualitative behavior of both measures is similar. Therefore, initially $S_l(\rho_A)$ and $C^c_{hs}(\rho_{AB})$ are equal (for p = 0), but, as the system evolves, $S_l(\rho_A) \neq C^c_{hs}(\rho_{AB})$, and the entanglement entropy is redistributed for the whole system.

\subsection{Correlated amplitude damping channel}
\label{subsec:correamp}
The correlated amplitude damping channel (CADC) was first considered in \cite{Yeo}:
\begin{equation}
    \rho \to \varepsilon(\rho) = (1 - \mu) \varepsilon_0 (\rho) + \mu \varepsilon_m (\rho),
\end{equation}
where $\mu \in [0,1]$ is the memory parameter. The memoryless ADC is recovered when $\mu = 0$ such that $\varepsilon_0$ represents the ADC, while for $\mu = 1$ we obtain the full memory amplitude damping channel $\varepsilon_m$. For $\varepsilon_m$ the relaxation phenomena are fully correlated, i.e., when a qubit undergoes a relaxation process, the other qubit does the same. Hence, the only state that has a probability different than zero of decaying is $\ket{1,1}_{A,B}$, while the other states are undisturbed. 

For $\mu=1$, the evolution of the whole system can be described by the following unitary map:
\begin{align}
    & U\ket{0,0}_{A,B}\ket{0,0}_{E_{A,B}} := \ket{0,0}_{A,B}\ket{0,0}_{E_{A,B}},\\
    & U\ket{1,0}_{A,B}\ket{0,0}_{E_{A,B}} := \ket{1,0}_{A,B}\ket{0,0}_{E_{A,B}},\\
    & U\ket{0,1}_{A,B}\ket{0,0}_{E_{A,B}} := \ket{0,1}_{A,B}\ket{0,0}_{E_{A,B}},\\
    & U\ket{1,1}_{A,B}\ket{0,0}_{E_{A,B}} := \sqrt{1 - p}\ket{1,1}_{A,B}\ket{0,0}_{E_{A,B}} + \sqrt{p}\ket{0,0}_{A,B}\ket{1,1}_{E_{A,B}}.
\end{align}
Here we denoted the environment as $E_{A,B}$ since, in this case, the environment is global, as discussed before. As we will see, this channel will generate simultaneous correlation between subsystems A and B and the environment. Hence,  if the initial state of the bipartite quantum system $AB$ is given by $\ket{\Psi}_{A,B} = x \ket{0,0}_{A,B} + \sqrt{1 - x^2}\ket{1,1}_{A,B}$ as before, then, after the coupling with the environment, the evolved global state is a pure state given by
\begin{align}
    \ket{\Psi}_{A,B,E_{AB}} & := U \ket{\Psi}_{A,B} \otimes \ket{0,0}_{E_A,E_B} \nonumber\\
    & = x\ket{0,0}\ket{0,0} + \sqrt{1 - x^2}\Big(\sqrt{1-p}\ket{1,1}\ket{0,0} + \sqrt{p}(\ket{0,0}\ket{1,1}\Big) \label{eq:camp},
\end{align}
meanwhile the evolved density operator of the partition AB, obtained by tracing out the degrees of freedom of the reservoirs, reads
\begin{equation}
\rho_{AB} =  \begin{pmatrix}
x^2 + p(1 - x^2) & 0 & 0 & \sqrt{1-p} x\sqrt{1 - x^2}\\
0 & 0 & 0 & 0\\
0 & 0 & 0 & 0\\
\sqrt{1-p} x\sqrt{1 - x^2} & 0 & 0 & (1 - p)(1 - x^2)\\
\end{pmatrix}.
\end{equation}

\begin{figure}[t]
\subfigure[\footnotesize Correlation measures as a function of p for the state $x = 1/4$.]{\includegraphics[scale = 0.45]{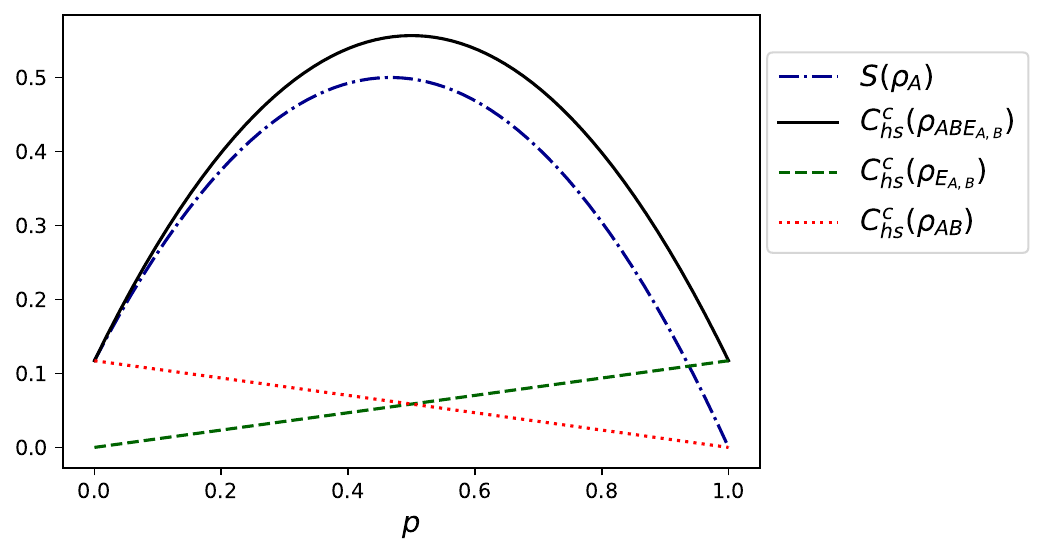} \label{fig:camp_}}
\subfigure[\footnotesize Correlation measures as a function of p for the state $x = 1/\sqrt{2}$.]{\includegraphics[scale = 0.45]{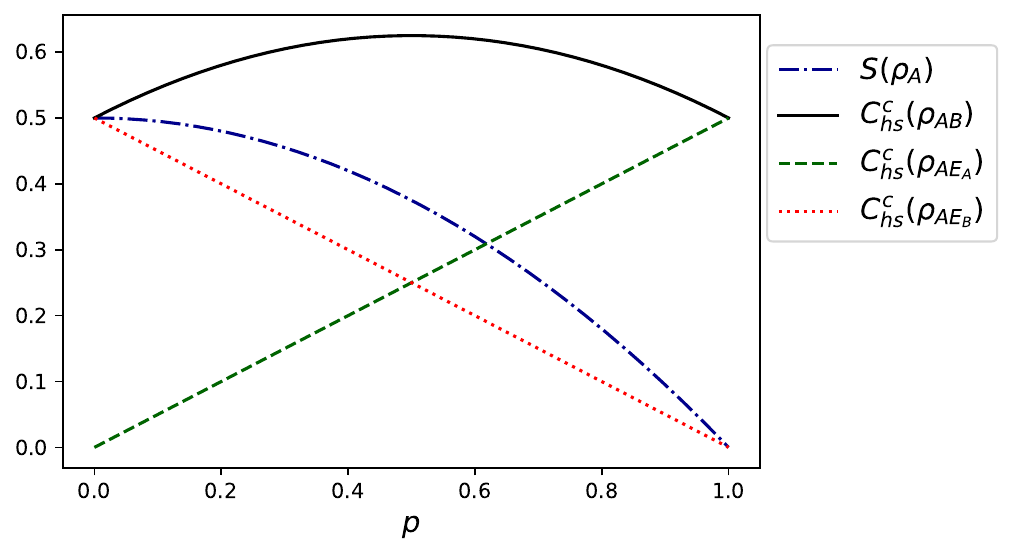} \label{fig:camp2}}
\qquad
\subfigure[\footnotesize Comparison between $C^c_{re}$ and $C^c_{hs}$ as a function of $p$ for $x = 1/\sqrt{2}$.]{\includegraphics[scale = 0.45]{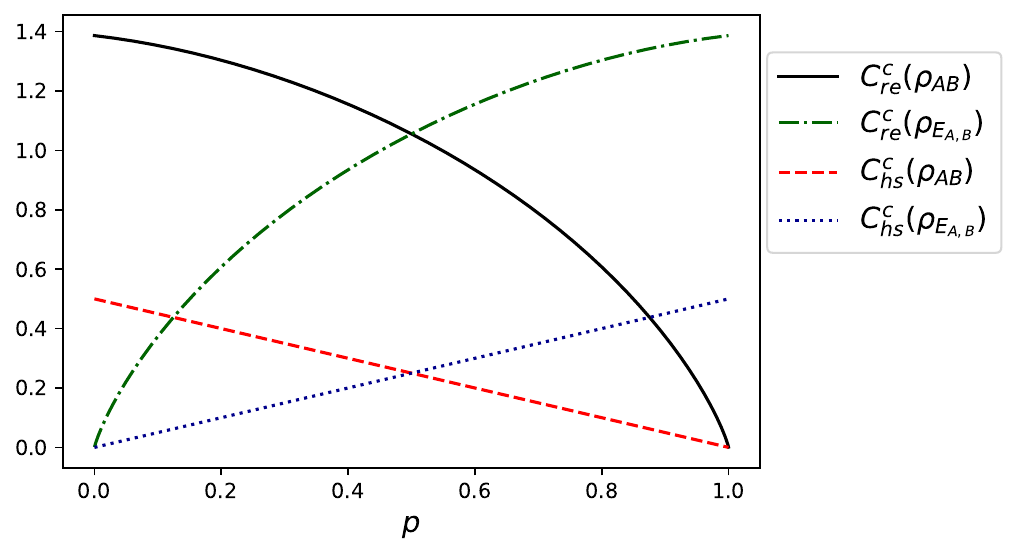}\label{fig:coecorr1}}
\caption{Entanglement entropy of $A$ and correlated quantum coherence of the different bipartitions as a function of $p$ for the correlated amplitude damping channel.}
\label{fig:camp}
\end{figure}

As we can see, the off diagonal elements of $\rho_{AB}$ are affected by a factor $\sqrt{1 - p}$ and, by the PPT criterion, the state $\rho_{AB}$ is entangled $\forall p \in [0,1)$ and $x \in (0,1)$. However $S_l(\rho_A) = 1 - (x^2 + (1 - x^2)p)^2 - (1 - p)^2(1 -x^2)^2$ and $C^c_{hs}(\rho_{AB}) = \frac{1}{2}\mathcal{C}^2_{AB} = 2(1 - p)x^2(1 - x^2)$, such that $S_l(\rho_A) > C^c_{hs}(\rho_{AB})$, which means that the initial entanglement between $A$ and $B$ is redistributed for the rest of the system. Hence, the correlation between $A$ and $B$ together with $P_{hs}(\rho_A)$ are not enough to completely characterize the quanton $A$. Since the environment is global, if we consider the bipartition $A E_A$, we obtain an incoherent state that is not entangled, since by the PPT criterion $(\rho_{A E_A})^{T_k} = \rho_{A E_A}$, where $\rho^{T_k}$ is the partial transposition of the density matrix $\rho$ with respect to the subsystem $k = A, E_A$. Therefore, it is necessary to consider the partition $A E_{A,B}$:
\begin{align}
    \rho_{A E_{A,B}} & = x^2 \ketbra{0,0,0} + (1 - p)(1 - x^2) \ketbra{1,0,0} + p(1 - x^2) \ketbra{0,1,1} \nonumber \\
    & + (\sqrt{p}x\sqrt{1 - x^2}\ketbra{0,0,0}{0,1,1} + t.c.),
\end{align}
where $t.c.$ stands for the transpose conjugate. However, the correlated coherence of this state is due only to the superposition of the states of the environments, as one can see by looking at the last two terms of the density operator. This means that we also have to consider the correlations in whole system $ABE_{A,B}$. By taking the partial trace of the subsystem $A$, we can see that $\rho_{E_{A,B}}$ is an entangled state
\begin{equation}
    \rho_{E_{A,B}} = \begin{pmatrix}
x^2 + (1 - p)(1 - x^2) & 0 & 0 & \sqrt{p}x\sqrt{1 - x^2}\\
0 & 0 & 0 & 0\\
0 & 0 & 0 & 0\\
\sqrt{p}x\sqrt{1 - x^2} & 0 & 0 & p(1 - x^2)\\
\end{pmatrix} .
\end{equation}
By noticing that the multipartite quantum coherence and the correlated quantum coherence are equal and measures the correlated coherences for the partitions $AB$, $ABE_{A,B}$, and $E_{A,B}$, we see that we have to discard the correlated coherence of the environment in order to quantify the correlations between $A$ and the rest of the system, since the correlation of the environment does not matter for the complete complementarity relation of the subsystem $A$. Thus
\begin{equation}
    C^c_{hs}(\rho_{A B E_{A,B}}) = 2x^2(1 - x^2)(1 - p) + 2x^2(1 - x^2)p + 2(1-x^2)^2(1-p)p,
\end{equation}
which is easily calculated from Eq. (\ref{eq:camp}). Another reason to discard the correlated coherence of the environment is that $C^c_{hs}(\rho_{A B E_{A,B}})$ is greater than $S_l(\rho_A)$ for some cases, as one can see from Fig. \ref{fig:camp}. Meanwhile, the correlated coherence of the environment is given by $C^c_{hs}(\rho_{ E_{A,B}}) = \frac{1}{2} \mathcal{C}^2_{E_{A,B}} = 2x^2(1 - x^2)p$, and therefore it is straightforward to show that
\begin{equation}
S_l(\rho_A) = C^c_{hs}(\rho_{A B E_{A,B}}) - C^c_{hs}(\rho_{E_{A,B}}),
\end{equation}
yielding the following complementarity relation for the subsystem $A$:
\begin{equation}
    P_{hs}(\rho_A) + C^c_{hs}(\rho_{A B E_{A,B}}) - C^c_{hs}(\rho_{ E_{A,B}}) = 1/2.
\end{equation}
Besides that, from Figs. \ref{fig:camp_}, \ref{fig:camp2}, and \ref{fig:coecorr1}, it is straightforward to see that $C^c_{hs}(\rho_{E_{A,B}})$ and $C^c_{hs}(\rho_{AB})$ are complementary to each other and obey the following complementarity relation
\begin{equation}
    C^c_{hs}(\rho_{E_{A,B}}) + C^c_{hs}(\rho_{AB}) = 2x^2(1-x^2).
\end{equation}
Therefore, one can see that $S_l(\rho_A)$ and $C^c_{hs}(\rho_{AB})$ are equal initially (for $p = 0$), but, as the system evolves, $S_l(\rho_A) \neq C^c_{hs}(\rho_{AB})$ and the entanglement entropy is redistributed for the whole system such that $S_l(\rho_A) = C^c_{hs}(\rho_{A B E_{A,B}}) - C^c_{hs}(\rho_{E_{A,B}})$. On the other hand, $C^c_{hs}(\rho_{AB})$ is turned into $C^c_{hs}(\rho_{E_{AB}})$. In the asymptotic time $p \to 1$, all the correlations are transferred to the environment. 

In Fig. \ref{fig:coecorr1}, we compare the qualitative behavior of $C^c_{hs}$ and the basis-independent measure $C^c_{re}$ to emphasize that $C^c_{hs}$ is invariant under local unitary transformations on the environments, with the additional condition that the local quantum coherences of $E_{A,B}$ remains null. It is interesting to note as well that the $C^c_{re}(\rho_{AB})$ and $C^c_{re}(\rho_{E_{A,B}})$ are complementary to each other.

\subsection{Phase damping channel}
\label{subsec:phasedam}
The phase damping channel (PDC) describes the loss of quantum coherence without loss of energy. It leads to decoherence without relaxation. An example of a physical system described by this channel is the random scattering of a photon by a heavy particle. Hence, this channel does not change the system base states, i.e., it does not exchange energy with the environment, since there is no transition between the states of the system. However, the system leaves a unique ``fingerprint'' in the environment, causing it to conditionally jump to a different configuration. The evolution of the whole system can be described by the following unitary map
\begin{equation}
    U\ket{j,0}_{k, E_k} := \delta_{j,0} \ket{0,0}_{k, E_k} + \delta_{j,1}(\sqrt{1 - p}\ket{0,0}_{k, E_k} + \sqrt{p} \ket{1,1}_{k, E_k}),
\end{equation}
with $j = 0,1$ and $k = A,B$. In this case, each subsystem has its own environment and the probability of each one to affect its local environment is the same. Now, if the initial state of the bipartite quantum system $AB$ is given by $\ket{\Psi}_{A,B} = x \ket{0,0}_{A,B} + \sqrt{1 - x^2}\ket{1,1}_{A,B}$, with $x \in [0,1]$, the evolved joint state of the system plus environment is given by
\begin{align}
\ket{\Psi}_{A,B,E_A,E_B} & = \Big(x \ket{0,0} + (1 - p)\sqrt{1 - x^2}\ket{1,1}\Big) \ket{0,0} + \sqrt{1 - x^2}\sqrt{p(1 - p)} (\ket{1,1}\ket{1,0} + \ket{1,1}\ket{0,1}) \nonumber \\
& + \sqrt{1 - x^2}p \ket{1,1}\ket{1,1} \label{eq:damp},
\end{align}
meanwhile the evolved density operator of the partition $AB$, obtained by tracing out the degrees of freedom of the reservoirs, is given by 
\begin{equation}
\rho_{AB} =  \begin{pmatrix}
x^2 & 0 & 0 & (1 - p) x\sqrt{1 - x^2}\\
0 & 0 & 0 & 0\\
0 & 0 & 0 & 0\\
(1-p) x\sqrt{1 - x^2} & 0 & 0 & (1 - x^2)\\
\end{pmatrix}.
\end{equation}
As we can see, in this case the diagonal elements of $\rho_{AB}$ are not affected by the interaction with the environment, which implies that the previsibility, $P_{hs}(\rho_k)$, and the linear entropy, $S_l(\rho_k)$, for $k = A,B$, are also not affected, since the individual states of $A$ and $B$ are incoherent states. While the off diagonal elements of $\rho_{AB}$ are affected by the factor $1 - p$, with $\rho_{AB}$ being an entangled state for $x \in (0,1)$ and $p \in [0,1)$, we have $S_l(\rho_A) > C^c_{hs}(\rho_{AB})$. Therefore, the entanglement of the two-qubit initial state was redistributed to the whole system.

\begin{figure}[t]
\subfigure[\footnotesize Correlation measures as a function of p for the state $x = 1/4$.]{\includegraphics[scale = 0.5]{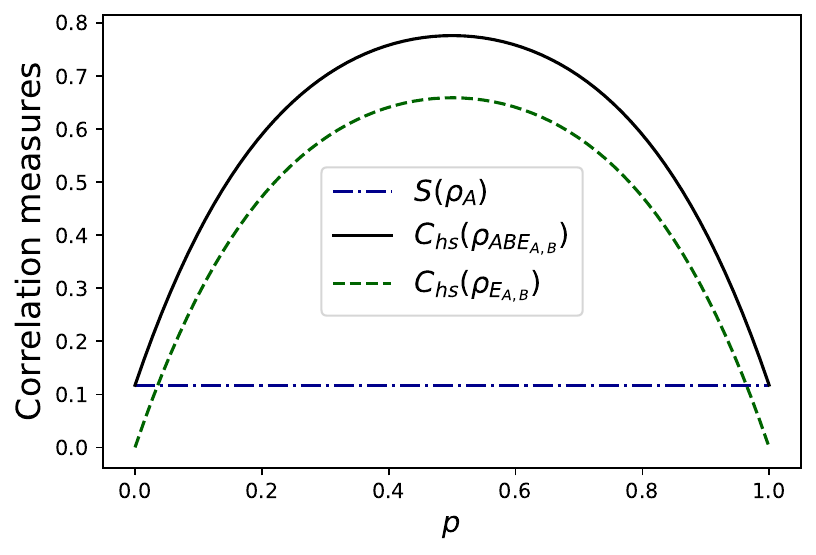} \label{fig:phase}}
\subfigure[\footnotesize Correlation measures as a function of p for the state $x = 1/\sqrt{2}$.]{\includegraphics[scale = 0.5]{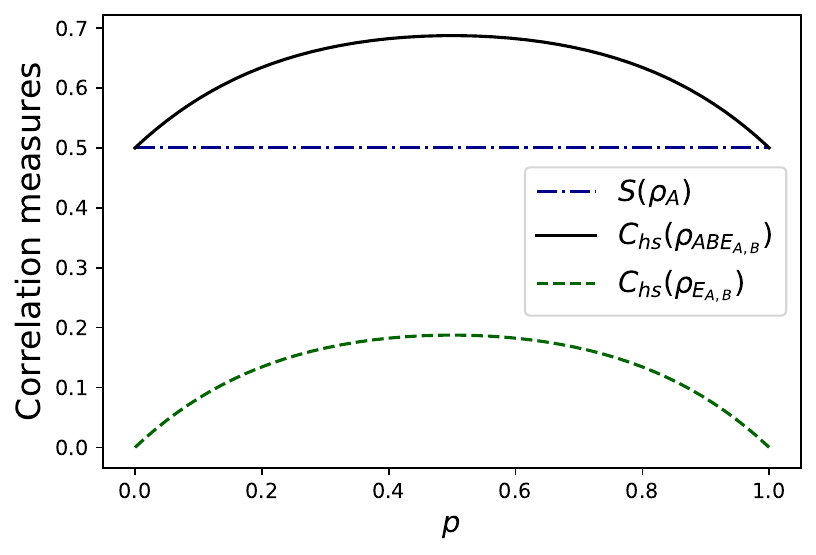} \label{fig:phase1}}
\subfigure[\footnotesize $C^{nl}_{hs}(\rho_{ABE_A}) + C^{nl}_{hs}(\rho_{ABE_B}) + C^{nl}_{hs}(\rho_{ABE_AE_B})$ as a function of $p$ for different values of $x$.]{\includegraphics[scale = 0.5]{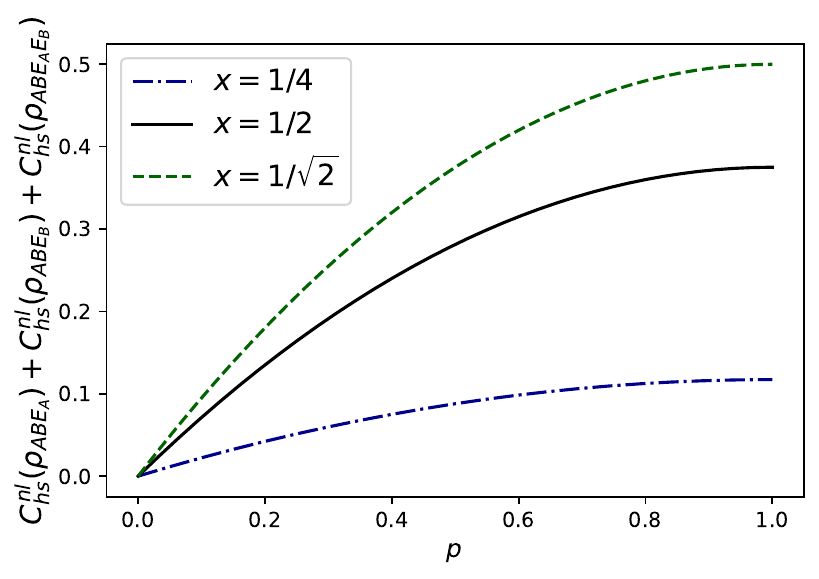} \label{fig:phase3}}
\subfigure[\footnotesize$C^{nl}_{hs}(\rho_{AB})$ as a function of $p$ for different values of $x$.]{\includegraphics[scale = 0.5]{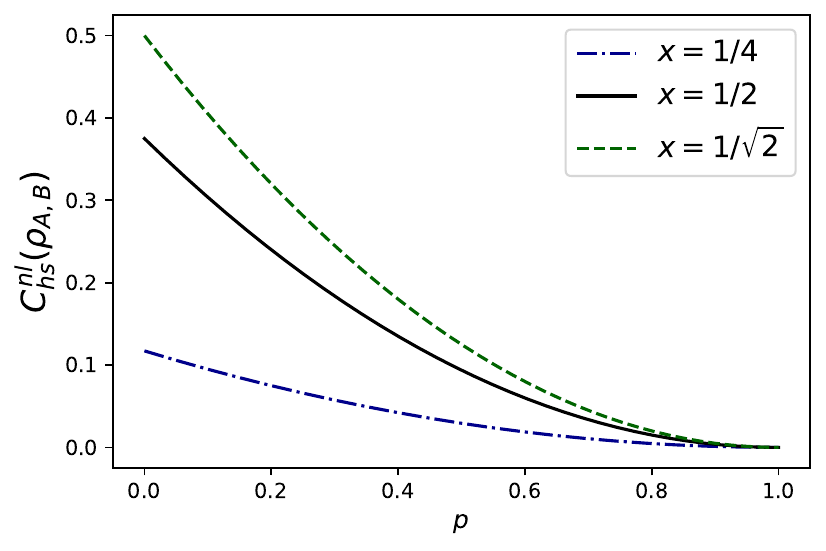} \label{fig:phase4}}
\caption{Entanglement entropy of $A$ and multipartite quantum coherence of the different partitions of the system as a function of $p$ for the phase damping channel.}
\label{fig:examp3}
\end{figure}

Now, we consider the state of the bipartitions $A E_A$, $A E_B$ and $E_A E_B$:
\begin{align}
    & \rho_{A E_A} = \rho_{A E_B} =  \begin{pmatrix}
x^2 & 0 & 0 & 0\\
0 & 0 & 0 & 0\\
0 & 0 & (1 - p)(1 - x^2) & \sqrt{p(1-p)}(1 - x^2)\\
0 & 0 & \sqrt{p(1-p)}(1 - x^2) & p(1 - x^2)\\
\end{pmatrix}, \\
 & \rho_{E_A E_B} =  \begin{pmatrix}
x^2 + (1 - p)^2(1 - x^2) & \gamma \sqrt{p(1-p)} & \gamma \sqrt{p(1-p)} & p (1 - p) (1 - x^2)\\
\gamma \sqrt{p(1-p)} & p(1 - p)(1 - x^2) & p (1 - p) (1 - x^2)& \zeta \sqrt{p(1-p)} \\
\gamma \sqrt{p(1-p)} & p (1 - p) (1 - x^2) & p(1 - p)(1 - x^2) & \zeta \sqrt{p(1-p)}\\
p(1 - p)(1 - x^2) & \zeta \sqrt{p(1-p)} & \zeta \sqrt{p(1-p)} & p^2(1 - x^2)\\
\end{pmatrix},
\end{align}
where $\gamma = (1 - x^2)(1 - p)$ and $\zeta = (1 - x^2)p$. By the PPT criterion, we  can directly see that  $(\rho_{A E_A})^{T_A} = \rho_{A E_A}$, $(\rho_{A E_B})^{T_A} = \rho_{A E_B}$, and $(\rho_{E_A E_B})^{T_{E_A}} = \rho_{E_A E_B}$, which means that there is no entanglement between the subsystems $A$ and $E_{A(B)}$ nor
between the reservoirs $E_A$ and $E_B$, for any value of $p$ and $x$ in $[0,1]$. Also, the coherences in $\rho_{AE_A}$ and $\rho_{AE_B}$ are due only to the environment. Although no bipartite entanglement is obtained beyond that contained in the two-qubit initial state, multipartite entanglement is always possible, as one can see by looking to Eq. (\ref{eq:damp}). In addition, by inspecting Eq.(\ref{eq:damp}), the coherences of such state are due to the partitions $AB$, $ABE_A$, $ABE_B$, $A B E_A E_B$, $E_A E_B$, $E_A$ and $E_B$:
\begin{align}
C_{hs}(\rho_{ABE_AE_B}) = & \underbrace{2x^2(1-x^2)(1-p)^2}_{AB} + \underbrace{4x^2(1-x^2)p(1-p)}_{ABE_A \ \& \ ABE_B}  + \underbrace{2x^2(1-x^2)p^2}_{ABE_AE_B} + \underbrace{4(1-x^2)^2(1 - p)^2p^2}_{E_AE_B} \nonumber \\
& + \underbrace{4(1-x^2)^2(1 - p)^3 p}_{E_A \ \& \ E_B} + \underbrace{4(1-x^2)^2(1 - p) p^3}_{E_A \ \& \ E_B},
\end{align}
meanwhile the bipartite coherence of the bipartition $E_A E_B$ is given by
\begin{equation}
C_{hs}(\rho_{E_A E_B}) = \underbrace{4(1-x^2)^2(1 - p)^2p^2}_{E_AE_B} + \underbrace{4(1-x^2)^2(1 - p)^3 p }_{E_A \ \& \ E_B} + \underbrace{4(1-x^2)^2(1 - p) p^3}_{E_A \ \& \ E_B}, 
\end{equation}
where the the underbrace of $E_A E_B$ means that the quantum coherence is stored globally, while $E_A \& E_B$ refers to the quantum coherence that is stored locally. By subtracting the last two equations, it follows that
\begin{align}
    C_{hs}(\rho_{ABE_AE_B}) - C_{hs}(\rho_{E_A E_B}) & =  \underbrace{2x^2(1-x^2)(1-p)^2}_{AB} + \underbrace{4x^2(1-x^2)p(1-p)}_{ABE_A \ \& \ ABE_B}  + \underbrace{2x^2(1-x^2)p^2}_{ABE_AE_B}\\
    & = S_l(\rho_A).
\end{align}
It is worth mentioning that the bipartite quantum coherences $C_{hs}(\rho_{E_A E_B})$ are not related with entanglement in this case. Besides, by defining the quantities $C^{nl}_{hs}(\rho_{AB}) := 2x^2(1-x)^2(1-p)^2$, $C^{nl}_{hs}(\rho_{ABE_A}) := C^{nl}_{hs}(\rho_{ABE_B}):= 2x^2(1-x^2)p(1-p)$, and $C^{nl}_{hs}(\rho_{ABE_AE_B}) := 2(1-x^2)^2x^2p^2$, which is not necessarily equal to the correlated quantum coherence. In fact, $C^{nl}_{hs}(\rho_{AB})$ is the part of the multipartite quantum coherence $C_{hs}(\rho_{ABE_AE_B})$ that is stored globally between $A$ and $B$; $C^{nl}_{hs}(\rho_{ABE_A})$ is the part of the quantum coherence $C_{hs}(\rho_{ABE_AE_B})$ that is stored only between $A B E_A$; and so on. We can see, as in Figs. \ref{fig:phase3} and \ref{fig:phase4}, that $C^{nl}_{hs}(\rho_{AB})$ is complementary to $C^{nl}_{hs}(\rho_{ABE_A}) + C^{nl}_{hs}(\rho_{ABE_B}) + C^{nl}_{hs}(\rho_{ABE_AE_B})$:
\begin{equation}
 C^{nl}_{hs}(\rho_{AB}) + C^{nl}_{hs}(\rho_{ABE_A}) + C^{nl}_{hs}(\rho_{ABE_B}) + C^{nl}_{hs}(\rho_{ABE_AE_B}) = S_l(x),
\end{equation}
where $S_l(x)$ is the linear entropy, which is a constant for each value of x. Therefore, we have
\begin{equation}
    P_{hs}(\rho_A) + C_{hs}(\rho_{ABE_AE_B}) - C_{hs}(\rho_{E_A E_B}) = 1/2.
\end{equation}
As always, for $p = 0$, $S_l(\rho_A) = C^c_{hs}(\rho_{AB})$, but, as the system evolves, $S_l(\rho_A) \neq C^c_{hs}(\rho_{AB})$, and the entanglement entropy is redistributed in the whole system and turned into general quantum correlations such that $S_l(\rho_A) = C_{hs}(\rho_{ABE_AE_B}) - C_{hs}(\rho_{E_A E_B})$. In this particular case, we can not ensure that $C_{hs}(\rho_{ABE_AE_B})$ and $C_{hs}(\rho_{E_A E_B})$ are invariant under local unitary transformations, but, certainly, the combination $C_{hs}(\rho_{ABE_AE_B}) - C_{hs}(\rho_{E_A E_B})$ is invariant, since it is equal to $S_l(\rho_A)$.

\subsection{Bit flip channel}
\label{subsec:bitplip}
\begin{figure}[t]
\subfigure[\footnotesize $S_l(\rho_A)$ and the Hilbert-Schmidt correlated coherences  as a function of $p$.]{\includegraphics[scale = 0.575]{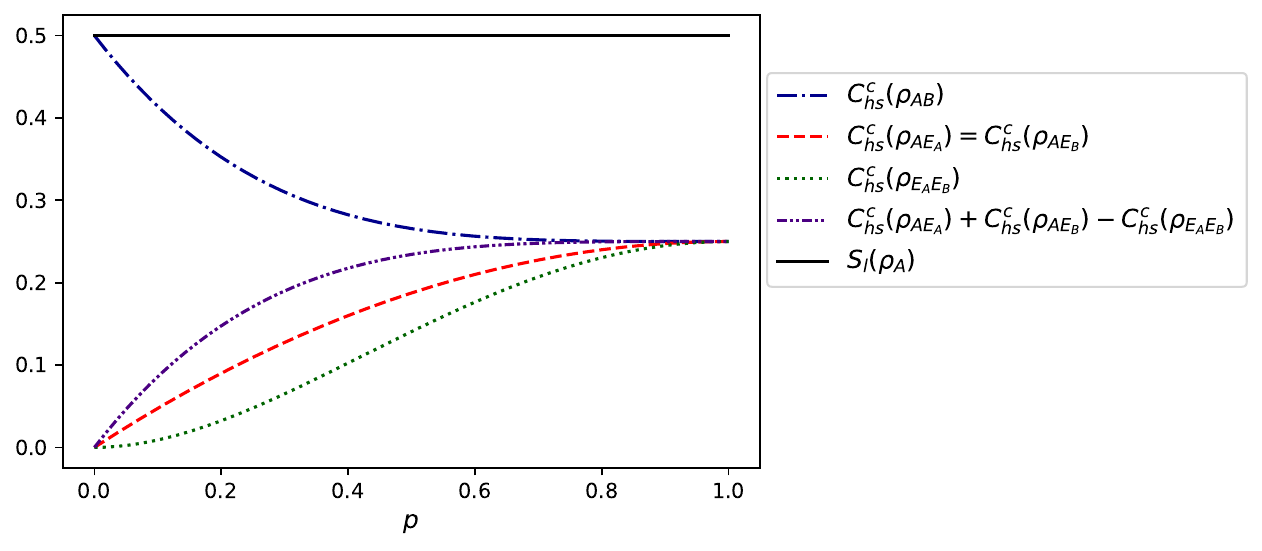} \label{fig:bitflip}}
\subfigure[\footnotesize  The relative entropy correlated coherences as a function of $p$.]{\includegraphics[scale = 0.575]{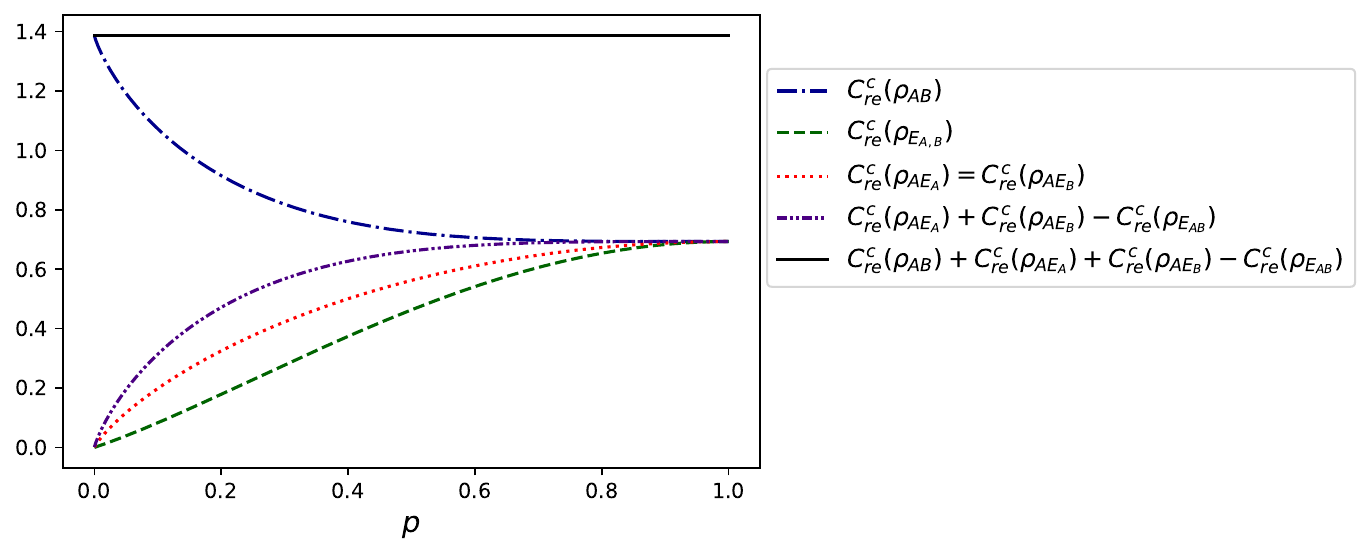} \label{fig:coecorr2}}
\caption{Correlation measures as a function of $p$ for the bit flip channel.}
\label{fig:examp2}
\end{figure}

The bit flip channel (BFC) is the most common error in classical information, where a bit can be flipped due to random noise. This type of error can also take place for a quantum bit, where the computational base states can be left alone $\ket{j} \to \ket{j}$, $j = 0,1$, with probability $1 - p/2$ or can be flipped, $\ket{j} \to \ket{k}$, $j,k = 0,1$, $j \neq k$, with  probability $p/2$. The unitary map that describes the BFC can be written as
\begin{align}
    U\ket{j,0}_{k,E_k} := \delta_{j,0}(\sqrt{1 - p/2}\ket{0,0}_{k,E_k} + \sqrt{p/2} \ket{1,1}_{k,E_k})  + \delta_{j,1}(\sqrt{1 - p/2}\ket{1,0}_{k,E_k} + \sqrt{p/2}\ket{1,1}_{k,E_k}),
\end{align}
with $j = 0,1$ and $k = A,B$. Now, if the initial state of the bipartite quantum system $AB$ is given by $\ket{\Phi^+}_{A,B} = \frac{1}{\sqrt{2}}(\ket{0,0}_{A,B} + \ket{1,1}_{A,B})$, the evolved joint state of the system plus environment is given by
\begin{align}
    \ket{\Phi}_{A,B,E_A,E_B} & = \frac{1}{\sqrt{2}}(1 - p/2)(\ket{0,0} + \ket{1,1})\ket{0,0} + \frac{1}{\sqrt{2}}\sqrt{(1 - p/2)p/2}(\ket{0,1} + \ket{1,0})(\ket{1,0} + \ket{0,1})\\
    & + \frac{1}{\sqrt{2}}p/2(\ket{0,0} + \ket{1,1})\ket{1,1},
\end{align}
where we choose $x = 1/\sqrt{2}$ for convenience, since there are more terms than for the previous channels, but the analysis is similar. The evolved density operator of the partition $AB$ is given by 
\begin{equation}
    \rho_{AB} = \frac{1}{4} \begin{pmatrix}
1 + (1 - p)^2 & 0 & 0 & 1 + (1 - p)^2\\
0 & 1 - (1 - p)^2 & 1 - (1 - p)^2 & 0\\
0 & 1 - (1 - p)^2 & 1 - (1 - p)^2 & 0\\
1 + (1 - p)^2 & 0 & 0 & 1 + (1 - p)^2\\
\end{pmatrix},
\end{equation}
which implies that $\rho_k = \frac{1}{2}I_{2 \times 2}$, where $I_{2 \times 2}$ is the $2\text{x}2$ identity matrix. Therefore, we can see that the diagonal elements of $\rho_{k}$ are not affected by the interaction with the environment, which implies that the predictability $P_{hs}(\rho_k)$, and the linear entropy $S_l(\rho_k)$, $k = A,B$ are also not affected. However, $S_l(\rho_A) > C^c_{hs}(\rho_{AB})$, which implies that the entanglement of the two-qubit initial state was redistributed in the whole system. Now, let us consider the state of the bipartitions $A E_A$, $A E_B$, and $E_A E_B$:
\begin{align}
    & \rho_{A E_A} = \rho_{A E_B} = \frac{1}{2} \begin{pmatrix}
1 - p/2 & 0 & 0 & \sqrt{(1 - p/2)p/2}\\
0 & p/2 & \sqrt{(1 - p/2)p/2} & 0\\
0 & \sqrt{(1 - p/2)p/2} & 1 - p/2 & 0\\
\sqrt{(1 - p/2)p/2} & 0 & 0 & p/2\\
\end{pmatrix}, \\
 & \rho_{E_A E_B} = \begin{pmatrix}
(1 - p/2)^2 & 0 & 0 & (1 - p/2)p/2\\
0 & (1 - p/2)p/2 & (1 - p/2)p/2 & 0\\
0 & (1 - p/2)p/2 & (1 - p/2)p/2 & 0\\
(1 - p/2)p/2 & 0 & 0 & p^2/4\\
\end{pmatrix}.
\end{align}
It follows directly that  $(\rho_{A E_A})^{T_A} = \rho_{A E_A}$, $(\rho_{A E_B})^{T_A} = \rho_{A E_B}$, and $(\rho_{E_A E_B})^{T_{E_A}} = \rho_{E_A E_B}$, which means that, for all values of $p$, there is no entanglement between the subsystems $A$ and $E_{A(B)}$ nor
between the reservoirs $E_A$ and $E_B$. From the reduced density matrices of all bipartitions, we can see that $\rho_{A}, \rho_{B}, \rho_{E_A}, \rho_{E_B}$ are all incoherent quantum states, which implies that the correlated quantum coherence and the bipartite quantum coherence are equal and given by
\begin{align}
    & C^c_{hs}(\rho_{AB}) = \frac{1}{4}(1 + (1 - p)^4),\\
    & C^c_{hs}(\rho_{AE_A}) = C^c_{hs}(\rho_{A E_B}) = (1 - p/2)p/2,\\
    & C^c_{hs}(\rho_{E_A E_B}) = 4(1 - p/2)^2(p/2)^2.
\end{align}
It is worth pointing out that, in this case, the correlated coherences $C^c_{hs}(\rho_{AE_A})$, $C^c_{hs}(\rho_{A E_B})$, and $C^c_{hs}(\rho_{E_A E_B})$ are not related to entanglement. Nevertheless, it's straightforward showing that
\begin{equation}
    S_l(\rho_A) = C^c_{hs}(\rho_{AB}) + C^c_{hs}(\rho_{AE_A}) + C^c_{hs}(\rho_{A E_B}) - C^c_{hs}(\rho_{E_A E_B}).
\end{equation}
So, one can see that $C^c_{hs}(\rho_{AB})$ is complementary to $C^c_{hs}(\rho_{AE_A}) + C^c_{hs}(\rho_{A E_B}) - C^c_{hs}(\rho_{E_A E_B})$, which also can be observed in Fig. \ref{fig:bitflip}. In Fig. \ref{fig:coecorr2} we plotted the behavior of the different basis-independent relative entropy of correlated coherence for comparison. For instance $C^c_{re}(\rho_{AB}) + C^c_{re}(\rho_{AE_A}) + C^c_{re}(\rho_{A E_B}) - C^c_{re}(\rho_{E_A E_B})$ is constant, with $C^c_{re}(\rho_{AB})$ being complementary to $C^c_{re}(\rho_{AE_A}) + C^c_{re}(\rho_{A E_B}) - C^c_{re}(\rho_{E_A E_B})$. It is worthwhile mentioning that for the phase flip (PFC) and bit-phase flip (BPFC) channels, the main-related behavior is the same.

\subsection{Phase flip channel}
\label{subsec:phaseflip}
\begin{figure}[t]
\subfigure[\footnotesize Measures of coherence and correlations as a function of $p$ for $x = 0.1$.]{\includegraphics[width=8.75cm]{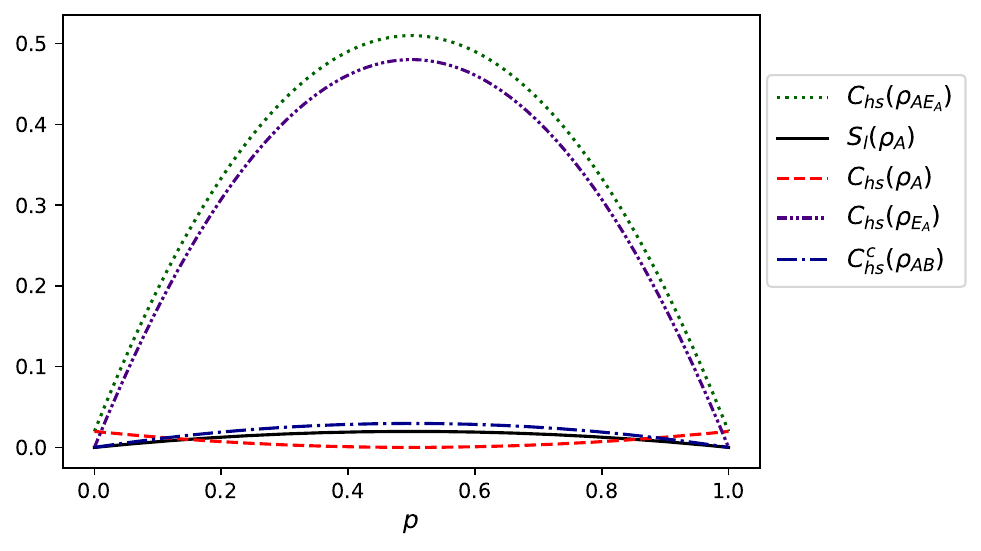} \label{fig:phaseflip3}}
\subfigure[\footnotesize Measures of coherence and correlations as a function of $p$ for $x = 0.5$.]{\includegraphics[width=8.75cm]{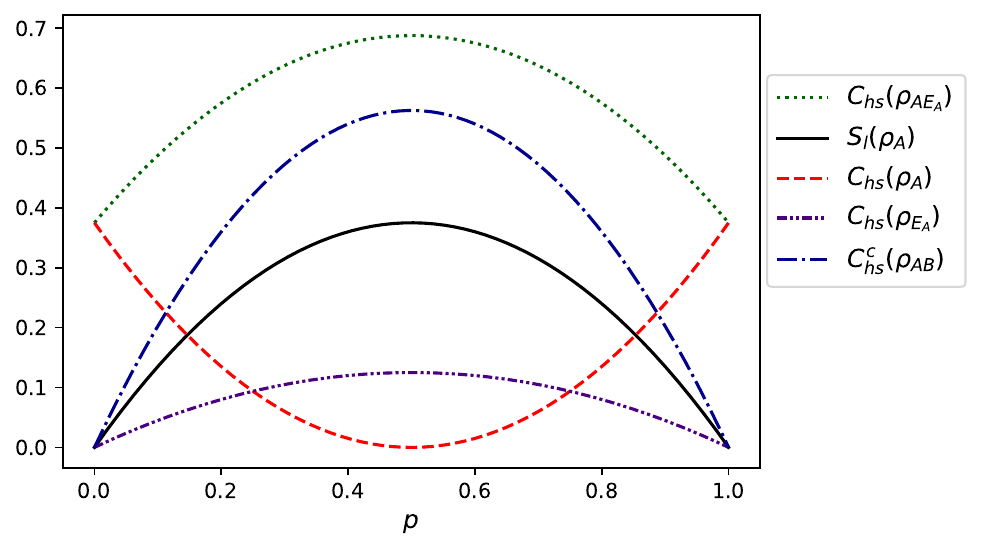} \label{fig:phasefli1}}
\subfigure[\footnotesize Measures of coherence and correlations as a function of $p$ for $x = 1/\sqrt{2}$.]{\includegraphics[width=8.75cm]{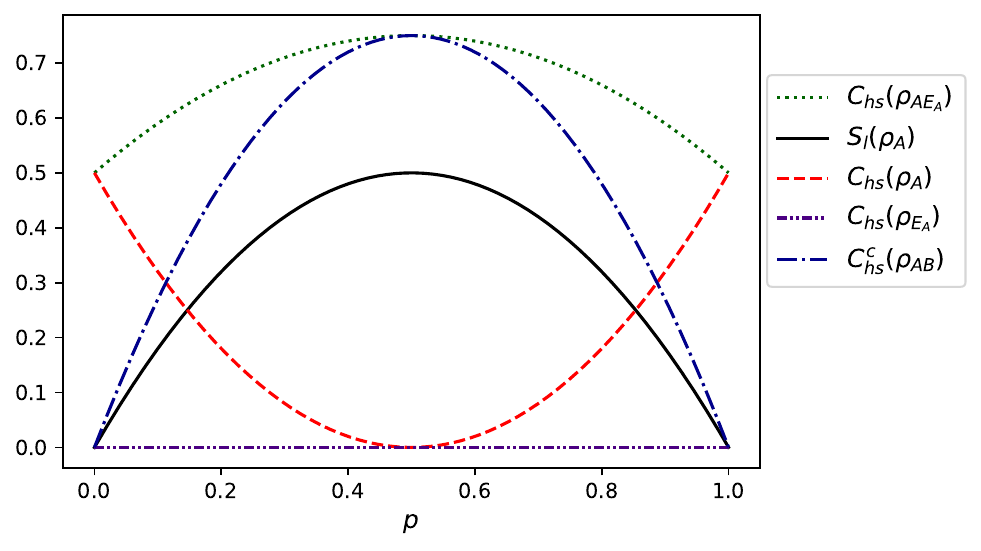}}
\caption{Measures of coherence and correlations as a function of $p$ for the phase flip channel.}
\label{fig:examp4}
\end{figure}

Since the analysis of the behavior of the entangled qubits under the phase flip channel (PFC) and under the bit-phase flip channel are similar to that for the bit flip channel, in the following sub-sections we'll study the behavior of a single qubit in a pure state interacting with its environment. The PFC is a kind of error which only happens in the quantum realm. Disregarding global phases, the state $\ket{0}_A$ is left unchanged with probability $1 - p$, whereas the state $\ket{1}_A$ acquires a $\pi$ phase, i.e., $\ket{1}_A \to e^{i\pi}\ket{1}_A$ with probability $p$, which leads to the following unitary map:
\begin{equation}
    U \ket{j,0}_{A,E_A} := \delta_{j,0}( \sqrt{1 - p}\ket{0,0}_{A,E_A} + \sqrt{p}\ket{0,1}_{A,E_A}) + \delta_{j,1}( \sqrt{1-p}\ket{1,0}_{A,E_A} - \sqrt{p}\ket{1,1}_{A,E_A}).
\end{equation}
Now, if the initial state of the qubit is given by $\ket{\psi}_A = x\ket{0}_A + \sqrt{1 - x^2}\ket{1}_A$, then the global state of the qubit and its environment will evolve to the following state
\begin{align}
    \ket{\psi}_{A,E_A} = x(\sqrt{1 - p}\ket{0,0} + \sqrt{p}\ket{0,1}) + \sqrt{1 - x^2}(\sqrt{1 - p}\ket{1,0} - \sqrt{p}\ket{1,1}), 
\end{align}
which in general is an entangled state. Meanwhile, the reduced states of $A$ and $E_A$ are given by
\begin{align}
    & \rho_A = x^2 \ketbra{0} + (1 - x^2)\ketbra{1} + \left((1 - 2p)x\sqrt{1-x^2}\ketbra{0}{1} + t.c.\right), \\
    & \rho_{E_A} = (1 - p) \ketbra{0} + p\ketbra{1} + \left(\sqrt{p(1 - p)}(2x^2 - 1)\ketbra{0}{1} + t.c.\right),
\end{align}
where $t.c.$ stands for the transpose conjugated. In general, it follows that $\rho_A$ is no longer a pure state. In addition, it is straightforward to see that the predictability is not affected by the coupling with the environment, since the diagonal elements of $\rho_A$ remain the same. On the other hand, the coherences of $\rho_A$ are affected by a factor $1 - 2p$, which implies that part of the quantum coherence of $\rho_A$ is turned into entanglement with the environment, since $S_l(\rho_A) \neq 0$ in general. If we denote $P_{hs}(\rho_A)_{p = 0} + C_{hs}(\rho_A)_{p = 0} = 1/2$ as the complementarity relation of the qubit $A$ before the coupling, then after the coupling we have the following complementarity relation: $P_{hs}(\rho_A) + C_{hs}(\rho_A) + S_l(\rho_A) = 1/2$. Since the predictability remains the same, it follows directly that $C_{hs}(\rho_A)_{p = 0} = C_{hs}(\rho_A) + S_l(\rho_A)$. However, in this case $S_l(\rho_A)$ is not equal to the correlated coherence of $AE_A$, even though they present the same dynamic behavior, as one can see in Fig. \ref{fig:examp4}. In addition, the measures of coherence and correlations of $A E_A$ are given by
\begin{align}
    & C_{hs}(\rho_{AE_A}) = 2(1 - p)p + 2x^2(1 - x^2)(1 - 2p + 2p^2),\\
    & C_{hs}(\rho_A) = 2(1 - 2p)^2x^2(1 - x^2),\\
    & C_{hs}(\rho_{E_A}) = 2p(1 - p)(2x^2 - 1)^2,\\
    & C^c_{hs}(\rho_{AE_A})=  C_{hs}(\rho_{AE_A}) - C_{hs}(\rho_A) -  C_{hs}(\rho_{E_A}),\\
    & S_l(\rho_A) = 1 - x^4 - (1 - x^2)^2 - 2(1 - 2p)^2x^2(1 - x^2).
\end{align}
From these equations, it follows directly that $C^c_{hs}(\rho_{AE_A}) = \frac{3}{2}S_l(\rho_A)$. Therefore $C^c_{hs}(\rho_{AE_A})$ is invariant under local unitary operations. Beyond that, we observe in Fig. \ref{fig:examp4} that, if the state of the qubit $A$ has low initial quantum coherence, the bipartite quantum coherence is almost entirely due to the quantum coherence of the environment $E_A$ as the system evolves. On the other hand, as $x \to 1/\sqrt{2}$ (i.e., as the initial state of the qubit $A$ approaches a maximal coherence state), the initial coherence of the quanton $A$ limits the maximum value of the quantum coherence of the environment $E_B$, while the correlated quantum coherence, $C^c_{hs}(\rho_{AE_A})$, and the entanglement entropy, $S_l(\rho_A)$, become more 
``distinguishable'' from each other, which is expected since their maximum values are different.

\subsection{Bit-phase flip channel}
\label{subsec:bitphase}

\begin{figure}[t]
\subfigure[\footnotesize Measures of the aspects of the qubit $A$ as a function of $p$ for $x = 0.1$.]{\includegraphics[width=8.75cm]{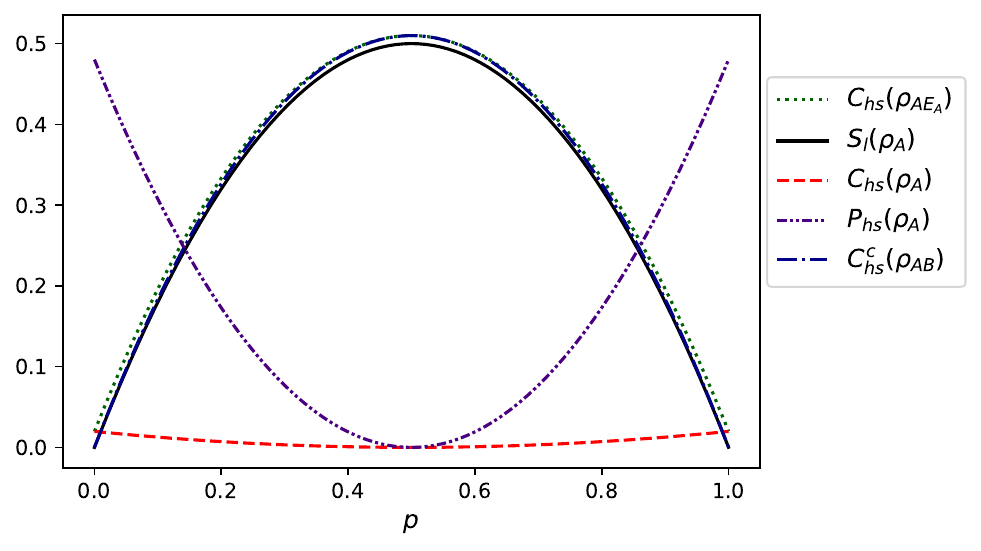} \label{fig:bitphase3}}
\subfigure[\footnotesize Measures of the aspects of the qubit $A$ as a function of $p$ for $x = 0.5$.]{\includegraphics[width=8.75cm]{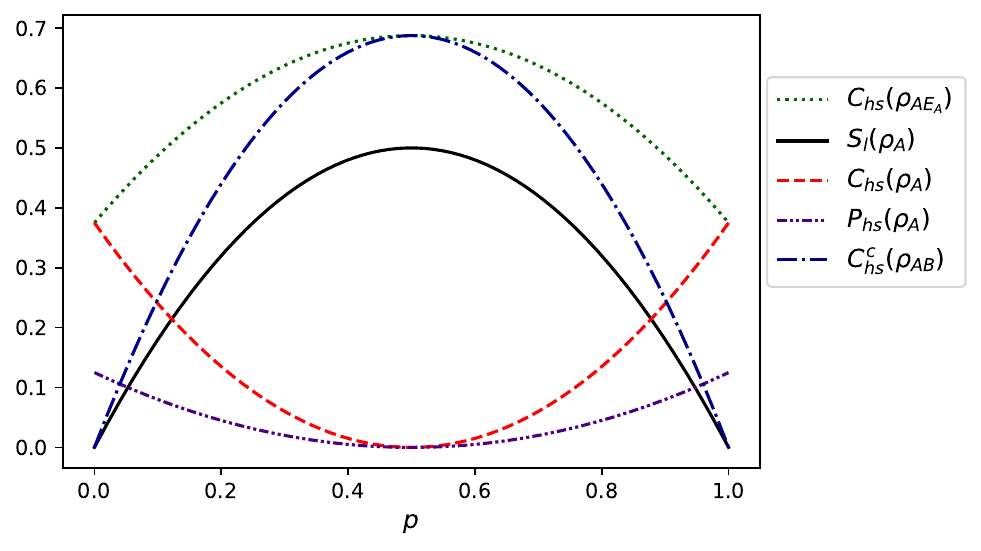} \label{fig:bitphase1}}
\subfigure[\footnotesize Measures of the aspects of the qubit $A$ as a function of $p$ for $x = 1/\sqrt{2}$.]{\includegraphics[width=8.75cm]{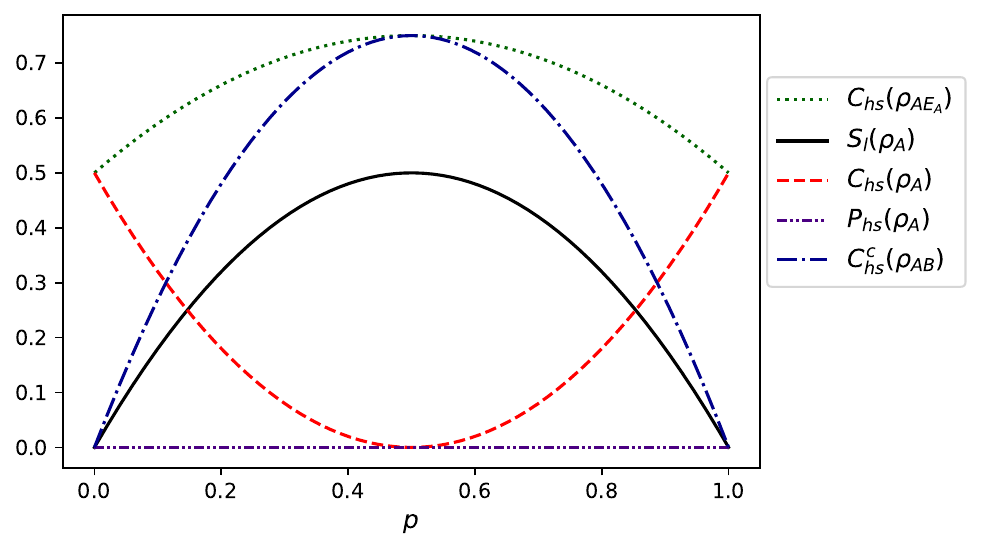}\label{fig:bitphase}}
\caption{Measures of the aspects of the qubit $A$ as a function of $p$ for the bit-phase flip channel.}
\label{fig:examp5}
\end{figure}

The bit-phase flip channel (BPFC) represents noises that are a mixture of those described by
phase and bit flip channels. Thus, the unitary map that describes the interaction between the system and the environment can be written as
\begin{equation}
U \ket{j,0}_{A,E_A} := \delta_{j,0}(\sqrt{1 - p}\ket{0,0}_{A,E_A} + i\sqrt{p}\ket{1,1}_{A,E_A}) + \delta_{j,1}( \sqrt{1-p}\ket{1,0}_{A,E_A} + -i \sqrt{p}\ket{0,1}_{A,E_A}).
\end{equation}
Then, if the initial state of the qubit is given by $\ket{\psi}_A = x\ket{0}_A + \sqrt{1 - x^2}\ket{1}_A$, the global state of the qubit and its environment will evolve to the following state
\begin{align}
    \ket{\psi}_{A,E_A} = x(\sqrt{1-p}\ket{0,0} + i\sqrt{p}\ket{1,1}) + \sqrt{1 - x^2}(\sqrt{1 - p}\ket{1,0} - i\sqrt{p}\ket{0,1}),
\end{align}
which in general is an entangled state. Meanwhile, the reduced states of $A$ and $E_A$ are given by
\begin{align}
    & \rho_A = (x^2 - p(1 - 2x^2)) \ketbra{0} + ((1 - x^2) - p(1-2x^2))\ketbra{1} + \left((1 - 2p)x\sqrt{1-x^2}\ketbra{0}{1} + t.c.\right), \\
    & \rho_{E_A} = (1 - p) \ketbra{0} + p\ketbra{1}.
\end{align}
In contrast to PFC, the BPFC affects the diagonal elements of $\rho_A$, which implies that in this case the predictability is also affected by the interaction with the environment. Therefore, it is not possible to conclude that part of the initial quantum coherence was turned into entanglement entropy. Meanwhile, $\rho_{E_A}$ is an incoherent state. Also, in this case, $S_l(\rho_A)$ is not equal to the correlated coherence of $AE_A$, even though they present the same behavior, as we can see in Fig. \ref{fig:examp5}. The different aspects of the qubit $A$ are given by 
\begin{align}
    & C_{hs}(\rho_{AE_A}) = 2x^2(1 - x^2) + 2(1 - p)p(x^4 + (1 - x^2)^2),\\
    & C_{hs}(\rho_A) = 2(1 - 2p)^2x^2(1 - x^2),\\
    & P_{hs}(\rho_{A}) = (x^2 + p(1 - 2x^2))^2 + ((1 -x^2) - p (1-2x^2))^2 - 1/2,\\
    & C^c_{hs}(\rho_{AE_A})=  C_{hs}(\rho_{AE_A}) - C_{hs}(\rho_A), \\
    & S_l(\rho_A) = 1/2 - P_{hs}(\rho_{A}) - C_{hs}(\rho_A),
\end{align}
from which it follows directly that $C^c_{hs}(\rho_{AE_A}) = \frac{3}{2}S_l(\rho_A)$.  In addition, if the state of the qubit A has low initial quantum coherence, as in Fig. \ref{fig:bitphase3}, the functions $C_{hs}(\rho_{AE_A})$, $C^c_{hs}(\rho_{AE_A})$ are very similar to each other. On the other hand, for $x \to 1/\sqrt{2}$, as in Figs. \ref{fig:bitphase1} and \ref{fig:bitphase}, the functions $C_{hs}(\rho_{AE_A})$, $C^c_{hs}(\rho_{AE_A})$, and $S_l(\rho_A)$ become more distinguishable from each other. Besides, it is interesting to notice that, as the state evolves ($p \to 1$), $P_{hs}(\rho_A)$ and $C_{hs}(\rho_A)$ decrease or increase together, although they reach different values. Therefore, in this case, the wave and particle aspects of the quanton A do not show a complementarity behavior between them, i.e., they reach the maximum and minimum possible values in the same point in the domain.

\subsection{Depolarizing channel}
\label{subsec:depo}
The depolarizing channel (DC) describes the situation in which the interaction of the system with the surroundings mixes its state with the maximally entropic one. We can  describe it by saying that, with probability $p$ the qubit remains intact, while with probability $1 - p$ the qubit state is turned into a completely mixed one. In contrast with the most common way of defining the unitary map that describes the DC \cite{nielsen}, we will define the unitary map as
\begin{equation}
    U \ket{\psi,0}_{A, E_A} = \sqrt{\frac{1 + p}{2}} I^A_{2 \times 2} \ket{\psi, 0}_{A,E_A} + \sqrt{\frac{1 - p}{2}} \sigma^A_y \ket{\psi, 1}_{A,E_A},
\end{equation}
where $I^A_{2 \times 2}$ is the identity matrix, $\sigma^A_y := -i\ketbra{0}{1} + i \ketbra{1}{0}$ is the one of the Pauli matrices, with both matrices acting on the Hilbert space of the qubit $A$ and with the coefficients of $\ket{\psi}_A$ being real numbers. Hence, the reduced states of qubit $A$ and of its environment are given by
\begin{align}
    & \rho_A = p \ketbra{\psi} + \frac{1-p}{2}I_{2 \times 2},\\
    & \rho_{E_A} = \frac{1+p}{2}\ketbra{0}{0} + \frac{1-p}{2}\ketbra{1}{1}.
\end{align}
Then, if the initial state of the qubit is given by $\ket{\psi}_A = x\ket{0}_A + \sqrt{1 - x^2}\ket{1}_A$, the different wave-particle and correlation aspects of the qubit $A$, after the coupling with the environment, are given by
\begin{figure}[t]
\subfigure[\footnotesize Measures of the aspects of the qubit $A$ as a function of $p$ for $x = 0.1$.]{\includegraphics[width=8.75cm]{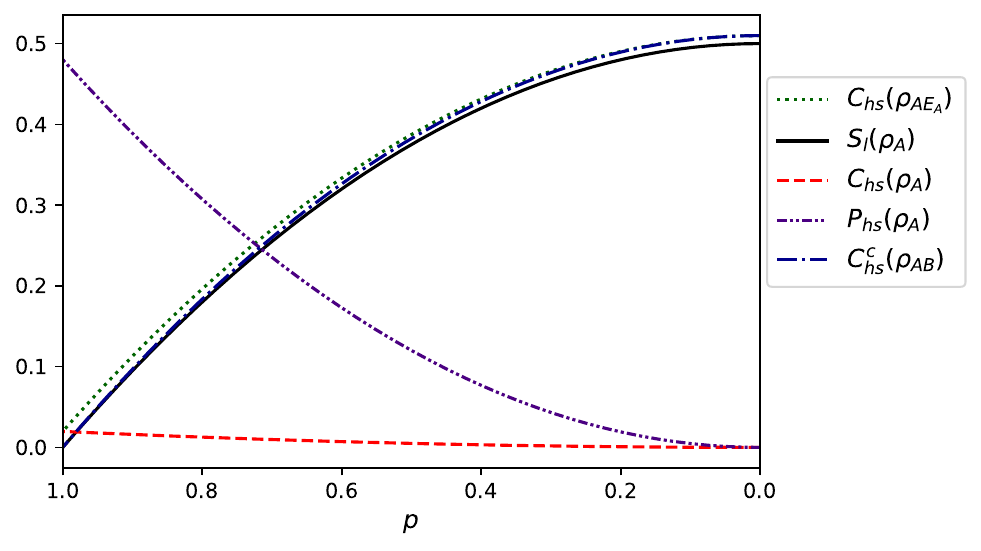} \label{fig:depo3}}
\subfigure[\footnotesize Measures of the aspects of the qubit $A$ as a function of $p$ for $x = 0.5$.]{\includegraphics[width=8.75cm]{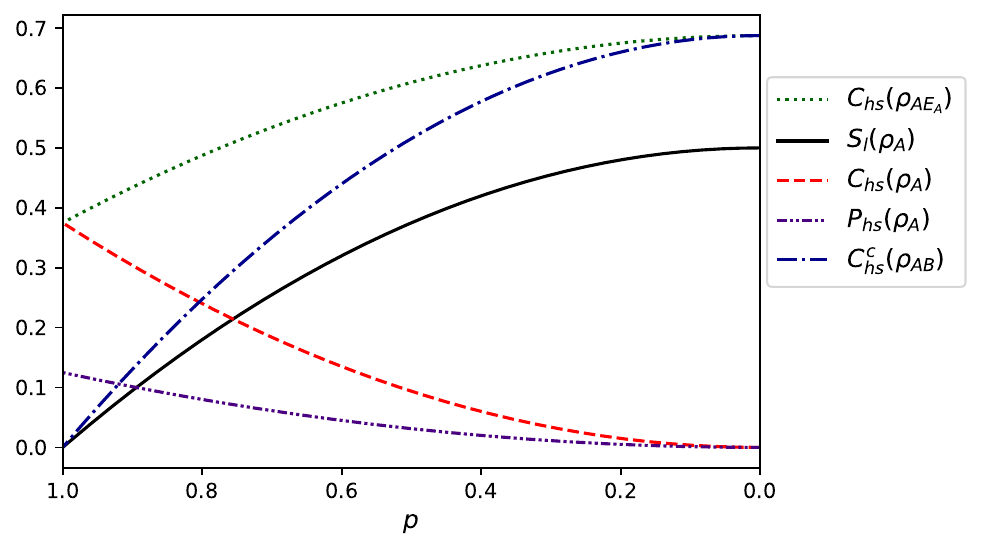} \label{fig:depo1}}
\subfigure[\footnotesize Measures of the aspects of the qubit $A$ as a function of $p$ for $x = 1/\sqrt{2}$.]{\includegraphics[width=8.75cm]{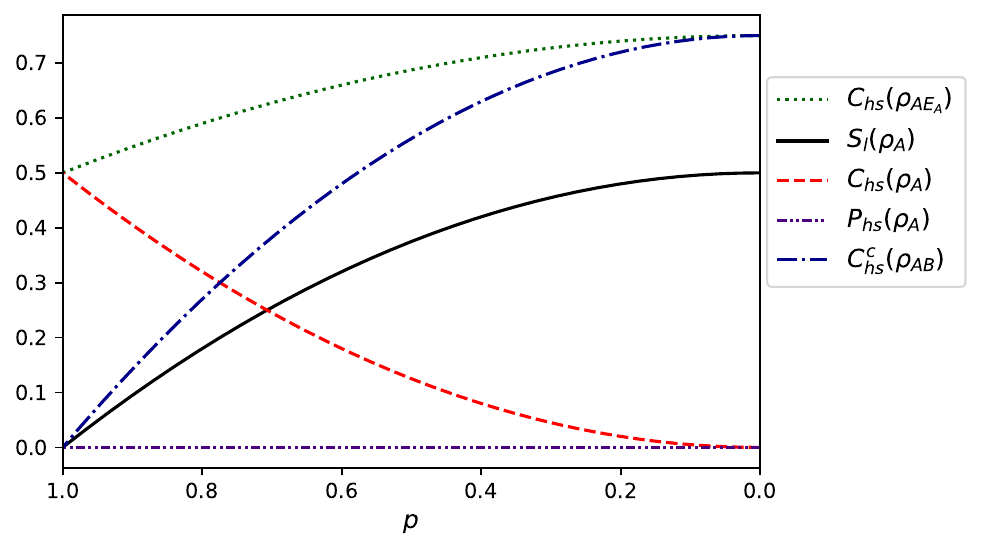}}
\caption{Measures of the wave-particle and correlation aspects of the qubit $A$ as a function of $p$ for the depolarizing channel.}
\label{fig:examp6}
\end{figure}
\begin{align}
    & C_{hs}(\rho_{A,E_A}) = (1 + p^2)x^2(1 - x^2) + \frac{1}{2}(1 - p^2),\\
    & C_{hs}(\rho_{A}) = 2p^2x^2(1 - x^2),\\
    & P_{hs}(\rho_A) = ( (1 - p)/2 + px^2)^2 + ( (1 + p)/2 + p(1 - x^2))^2 - 1/2,\\
    & C^c_{hs}(\rho_{A,E_A}) = C_{hs}(\rho_{A,B}) - C_{hs}(\rho_{A}),\\
    & S_l(\rho_A) = 1/2 - P_{hs}(\rho_A) - C_{hs}(\rho_{A}),
\end{align}
which implies that $C^c_{hs}(\rho_{A,E_A}) = \frac{3}{2}S_l(\rho_A)$. Once more, for $x \to 1/\sqrt{2}$ the functions $C^c_{hs}(\rho_{AE_A})$ and $S_l(\rho_A)$ diverge from each other, as shown in Fig. \ref{fig:examp6}. Also, we can see that for $p=1$, the aspects of the system are completely local.  However, for later times ($p \to 0$), the quantum coherence is stored globally, since $C_{hs}(\rho_{AE_A}) = C^c_{hs}(\rho_{AE_A})$ and $C_{hs}(\rho_A) = C_{hs}(\rho_{E_A}) = 0$. Even more interesting, both the predictability and the quantum coherence decrease together as $p \to 0$, except for the state given by $x = 1/\sqrt{2}$. Therefore, the local complementary aspects of the system are turned to non-local complementary aspects by the depolarizing channel.

\section{Conclusions}
\label{sec:conc}
Quantum channels are tools capable of providing a description for a wide range of physical situations. In this article, we used these tools to present a detailed study of complete complementarity relations of one and two qubits, belonging to a pure quantum system, under the interaction with environments modeled by some important quantum channels. 

By starting with an entangled bipartite pure quantum state, with the linear entropy being the quantifier of entanglement, we studied the dynamical flow of the linear entropy and how the entanglement entropy is redistributed and turned into correlations among the degrees of freedom of the whole system. Our investigation provided important insights about how these different kinds of system-environment interactions can affect the correlation (and complementarity) properties of bipartite quantum systems and of the environment, and about how they exchange these correlations. As we saw, it is always possible to express the entanglement entropy in terms of the multipartite quantum coherence, or in terms of the correlated quantum coherence, of the different partitions of the system, that in some cases quantifies entanglement but in others cases does not. For instance, the entanglement entropy was redistributed among the whole system for the amplitude damping channel such that $S_l(\rho_A) = C^{c}_{hs}(\rho_{AB}) +C^{c}_{hs}(\rho_{AE_A})  + C^{c}_{hs}(\rho_{AE_B})$, with $C^{c}_{hs}(\rho_{AB}), C^{c}_{hs}(\rho_{AE_A})$ being directly related to the concurrence measure of entanglement, while $C^{c}_{hs}(\rho_{AE_B})$ is not. On the other hand, for the phase damping channel, the entanglement entropy is redistributed among the whole system and turned to global quantum coherence such that $S_l(\rho_A) = C_{hs}(\rho_{ABE_AE_B}) - C_{hs}(\rho_{E_A E_B})$, where $C_{hs}(\rho_{ABE_AE_B})$ is measuring the total coherence of the different partitions of the entire system, and $C_{hs}(\rho_{E_A E_B})$
is the bipartite quantum coherence between $E_A$ and $E_B$.

In addition, for one qubit states under phase flip, bit-phase flip, and depolarizing channels, we showed that correlated quantum coherence and the entanglement entropy are directly related to each other, even though as the initial state of the qubit approaches a maximally coherent state, the correlated quantum coherence and the entanglement entropy become more `distinguishable' from each other. Even more interesting, we noticed that, for the depolarizing and bit-phase flip channels, the wave-particle aspects can decrease and increase together. This result provides examples of possible physical situations where complete complementarity relations are needed for a thorough quantification of the wave-particle aspects of a quanton. Besides, we showed that it is possible to consider the environment as part of the quantum system such that the whole system can be taken as a multipartite pure quantum system, which allowed us to consider linear entropy not just as a measure of mixedness of a particular subsystem, but as a correlation measure of the subsystem with rest of the universe.

\begin{acknowledgments}
This work was supported by the Coordena\c{c}\~ao de Aperfei\c{c}oamento de Pessoal de N\'ivel Superior (CAPES), process 88882.427924/2019-01, and by the Instituto Nacional de Ci\^encia e Tecnologia de Informa\c{c}\~ao Qu\^antica (INCT-IQ), process 465469/2014-0.
\end{acknowledgments}

\end{document}